\documentclass[12pt,a4paper]{article}
\pdfoutput = 1
\usepackage{jheppub}

\usepackage{lmodern} 
\usepackage[utf8]{inputenc} 
\usepackage[T1]{fontenc} 
\usepackage{microtype} 
\usepackage[english]{babel}
\usepackage{color}
\usepackage{multicol}
\usepackage{setspace}
\usepackage{amsmath,amsfonts,amssymb,dsfont}
\usepackage{mathrsfs}
\usepackage{stmaryrd}
\usepackage{slashed}

\usepackage{color}
\usepackage{empheq}
\usepackage{cancel}

\usepackage{graphicx}
\usepackage{tabularx}
\usepackage[usenames,dvipsnames]{xcolor}

\usepackage{tikz}
\usetikzlibrary{decorations.pathmorphing}

\newcommand{\erm}{\mathrm{e}}

\newcommand{\wrt}{with respect to }

\newcommand{\dd}{\mathrm{d}}
\newcommand{\e}{\mathrm{e}}
\newcommand{\eq}{\mathrm{eq}}
\newcommand{\half}{\frac{1}{2}}

\newcommand{\vev}[1]{\left\langle #1 \right\rangle}

\newcommand{\ii}{\textrm{i}}

\newcommand{\action}[1]{\mathcal{S}_\text{#1}}

\newcommand{\nwtn}[1]{\mathscr{G}_\text{N}^{(#1)}}

\newcommand{\h}{\text{h}}

\newcommand{\volt}{\mathcal{V}_{\mathbb{T}^2}}

\DeclareMathOperator{\arccot}{arccot}

\definecolor{myblue}{RGB}{0,0,125}
\definecolor{myred}{RGB}{125,0,0}
\definecolor{mypurple}{RGB}{125,0,125}
\definecolor{myviolet}{RGB}{125,0,250}
\definecolor{mygreen}{RGB}{0,125,0}
\definecolor{mycyan}{RGB}{0, 158, 115}
\definecolor{myorange}{RGB}{213,94,0}

\newenvironment{subeq}{
\subequations \align}
{\endalign \endsubequations
}

\begin{document}

\title{Schwarzian quantum corrections to shear correlators of the near-extremal Reissner-Nordstr\"om-AdS black hole}
\author[a]{Blaise Gout\'eraux,}
\author[a,b]{David M. Ramirez,}
\author[a]{and Cl\'ement Supiot}
\affiliation[a]{CPHT, CNRS, \'Ecole polytechnique, IP Paris, F-91128 Palaiseau, France}
\affiliation[b]{Department of Applied Mathematics and Theoretical Physics,
University of Cambridge, Cambridge CB3 0WA, UK}
\emailAdd{blaise.gouteraux@polytechnique.edu, david.ramirez@polytechnique.edu, clement.supiot@polytechnique.edu}
\preprint{
CPHT-R063.122025}
\abstract{Near-AdS$_2$ spacetimes are controlled by a Schwarzian effective dual theory. The Kaluza-Klein reduction of higher-dimensional black holes shows that the Schwarzian generates a logarithmic contribution to the entropy, thereby resolving a long-standing puzzle in near-extremal black hole thermodynamics. Here, we leverage exact results for quantum-corrected, Schwarzian scalar correlation functions  in order to evaluate the impact of bulk quantum fluctuations on the low-temperature shear correlators of the state dual to Reissner-Nordstr\"om-AdS$_4$ black holes with a flat, compact horizon. In the hydrodynamic regime, we find that quantum fluctuations tend to increase the shear viscosity away from $s/(4\pi)$, thereby preserving the Kovtun-Son-Starinets bound. Outside the hydrodynamic regime, quantum fluctuations lift the zero temperature, classical gapless modes reported in previous literature.}
\maketitle


\section{Introduction and summary of results}

At late times and long distances, interacting thermal systems are described by the theory of hydrodynamics \cite{forster2018hydrodynamic,chaikinlubensky1995,Kovtun:2012rj,Liu:2018kfw}. The starting point of this effective description is the notion that local thermal equilibrium is established over time and length scales $t\gtrsim\tau_{\eq}$, $x\gtrsim\ell_\eq$, or frequencies $\omega\lesssim\omega_\eq$ and wavenumbers $k\lesssim k_\eq$ in Fourier space. These local equilibrium scales mark the typical decay scales of non-conserved operators, i.e. those not protected by a global symmetry such as invariance under time or space translations. The relaxation from this collection of local equilibria to global equilibrium happens through hydrodynamic modes, such as diffusive or acoustic modes. In this regime, only the dynamics of conserved operators are retained through a set of conservation equations for hydrodynamic fields, e.g. energy, momentum or charge. The expectation values of non-conserved operators in the local equilibrium state are given by constitutive relations, which are expansions order by order in derivatives of the hydrodynamic fields. These derivative terms come with a finite number of transport coefficients, which must be computed in the original microscopic theory in order for the hydrodynamic theory to be predictive. Some of these transport coefficients are constrained by Onsager reciprocity (transformation rules under discrete symmetries like parity or time reversal) or by imposing the positivity of entropy production (the local second law of thermodynamics).

Another essential aspect is the discovery that the low-energy excitations of asymptotically anti de Sitter (AdS) black holes are simply those of hydrodynamics \cite{Policastro:2002se,Policastro:2002tn,Herzog:2002fn,Kovtun:2005ev,Bhattacharyya:2007vjd,Banerjee:2008th,Erdmenger:2008rm}. This is an enormous simplification of the gravitational theory, as it amounts to the statement that the complicated gravitational dynamics is those of a (charged) fluid
\begin{equation}
\nabla_\mu T^{\mu\nu}=0\,,\quad \nabla_\mu J^\mu=0\,,
\end{equation}
where the stress-energy tensor $T^{\mu\nu}$ and the U(1) current $J^\mu$ are evaluated in the local equilibrium state. 

\paragraph{Universality of shear viscosity in gauge/gravity duality}

One of the best-known transport coefficients is the shear viscosity $\eta$, which governs the diffusive relaxation of transverse momentum. In fact, in a neutral, conformal relativistic fluid, this is the only transport coefficient at first order in derivatives. As such, it plays a crucial role in modeling the dynamics of the Quark-Gluon-Plasma formed in heavy-ion collisions \cite{Berges:2020fwq}, which is approximately conformal and relativistic in the regime of interest. Entropy production only loosely constrains this transport coefficient, imposing that $\eta\geq0$.

Computing transport coefficients in a microscopic theory involves a Kubo formula and is typically very difficult, especially when elementary constituents are strongly-coupled. This highlights the importance of existing microscopic calculations of the shear viscosity, such as that in $\mathcal N=4$ super Yang-Mills theory using gauge/gravity duality \cite{Policastro:2001yc}. It is valid in the limit where the number of colors $N_\text{c}\to+\infty$ and where the `t Hooft coupling $\lambda = g_\text{YM}N_\text{c}^2\to+\infty$, for which the gauge theory has a gravitational dual in terms of classical Einstein gravity.

Remarkably, the shear viscosity in this regime is given by a very simple formula in terms of the entropy density $s$ of the black hole in the dual gravity theory,
\begin{equation}
\label{etas}
\eta=\frac{s}{4\pi}\,.
\end{equation}
The apparent simplicity of this result, and its universality in two-derivative Einstein gravity theories coupled to matter \cite{Kovtun:2003wp,Cremonini:2011iq}, motivated \cite{Kovtun:2004de,Son:2007vk}
to conjecture the bound \begin{equation}
\label{KSSbound}
\frac{\eta}{s}\gtrsim\frac1{4\pi}\,.
\end{equation}
The merit of such bounds cannot be understated. For any quantity which cannot be computed systematically given any microscopic theory, the ability to constrain low-energy observables through ideally tight bounds is tremendously useful in order to explore the parameter space of the corresponding low-energy effective field theory, see \cite{Hartnoll:2014lpa,Grozdanov:2020koi,Heller:2022ejw,Heller:2023jtd,Chowdhury:2025dlx,Chowdhury:2025qyc,Chowdhury:2025tzq} for other such examples. The bound \eqref{KSSbound} restricts the  parameter space allowed by the second law.

In fact, the relation \eqref{etas} and the bound \eqref{KSSbound} are very special to translation- and rotation-invariant states in two-derivative Einstein holographic theories. Including higher-derivative gravity terms spoil both \eqref{etas} and \eqref{KSSbound}, see \cite{Cremonini:2011iq} for a review, and similarly when breaking rotation or translation symmetries \cite{Rebhan:2011vd,Davison:2014lua,Jain:2014vka,Hartnoll:2016tri,Alberte:2016xja}. It is more plausible that diffusivities instead of transport coefficients  obey a fundamental bound \cite{Hartnoll:2014lpa}. In this work, we investigate the impact of gravitational quantum fluctuations from the near-extremal AdS$_2\times \mathbb{R}^2$ region of spacetime on \eqref{etas} and \eqref{KSSbound}. We will find that while \eqref{etas} no longer holds,  \eqref{KSSbound} still does.

\paragraph{Hydrodynamics and spectrum of collective excitations near extremality:}

For neutral black holes or charged black holes at temperatureslarge compared to the chemical potential $T\gtrsim\mu$, the local equilibrium scales $\omega_{\eq}$, $k_\eq$ were found to be of order the temperature \cite{Withers:2018srf,Grozdanov:2019kge,Grozdanov:2019uhi,Jansen:2020hfd}. Instead, for near-extremal black holes with an emergent AdS$_2\times \mathbb{R}^2$ near-horizon geometry, which have $T\ll\mu$ (equivalently, $Tr_\e\ll1$), while the equilibrium frequency remains $\omega_\eq\sim T$, the equilibrium wavenumber is enhanced to $k_\eq\sim\sqrt{T/r_\e}$ \cite{Arean:2020eus,Gouteraux:2025kta}, where $r_\e$ is the extremal radius. 

Surprisingly, gapless modes with $\omega(k\to0)=0$ survive in the extremal, $T=0$ limit of these planar black holes, with dispersion relations apparently given by the naive zero temperature limit of the corresponding hydrodynamic dispersion relations. For example, in the shear sector, a zero temperature mode was found $\omega=-\ii D_{\perp,0}k^2+O(k^4)$, where $D_{\perp,0}=\lim_{T\to0} D_\perp=s_0/(4\pi\mu\rho_0)$ \cite{Edalati:2009bi,Edalati:2010hk,Davison:2013bxa,Arean:2020eus}. Here, $s_0$ and $\rho_0$ are the entropy and charge densities at $T=0$. Similar modes are found in other sectors, such as the longitudinal sound, longitudinal thermal diffusion or charge diffusion sectors \cite{Edalati:2009bi,Edalati:2010pn,Gushterov:2018spg,Arean:2020eus,Gouteraux:2025kta}. From \eqref{etas}, it is tempting to invoke the fact that the entropy density of these black holes is non-vanishing in the extremal limit in order to explain the existence of the $T=0$ gapless modes. However, this argument only appears to apply directly to the shear sector where $D_{\perp,0}\sim s_0$, and not to other sectors.

These zero temperature gapless modes are problematic for two reasons. First, it should be clear that these are not hydrodynamic modes, as they arise on scales beyond the radius of convergence where hydrodynamics is a valid effective theory. In other systems where they occur, their existence is protected by an emergent global symmety at $T=0$. In Fermi liquids, zero sound is a consequence of the emergent loop U(1) group that follows from the conservation of quasiparticles around the Fermi surface \cite{Else_2021}. In $T=0$ superfluids, the conservation of winding causes superfluid sound \cite{Delacretaz:2019brr}. While the emergence of holographic zero temperature modes can sometimes explicitly be traced to an emergent higher-form symmetry \cite{Chen:2017dsy,Grozdanov:2018fic,Davison:2022vqh}, the physical origin of this symmetry is unclear. 

Second, these zero temperature holographic states also feature a different type of singularity in the complex frequency plane: branch cuts along the negative imaginary axis, which arise from the coalescence of a series of gapped poles $\omega_n=-2\pi \ii T(\Delta+n)$, for non-negative integers $n$, and where $\Delta$ is the scaling dimension of the least irrelevant operator in the IR theory \cite{Faulkner:2009wj,Edalati:2009bi,Edalati:2010hk,Edalati:2010pn,Hartnoll:2012rj}. Hence, the co-existence of a branch cut and a pole on the negative imaginary axis is not mathematically well-defined, unless these singularities add in series in the retarded Green's functions for some special reason. 

The second issue was resolved in the case of current-current retarded Green's functions in \cite{Gouteraux:2025kta}: computing the dispersion relation to order $O(k^4)$ reveals that it takes the form $\omega = - \ii r_\e k^2+\ii (2 r_\e^3/3)k^4(d_4\mp \ii\pi+2\log(kr_\e))+O(r_\e^5 k^6)$. This shows that in fact there is a pair of gapless poles with a nonzero real part stemming from non-analytic, logarithmic contributions caused by the branch cut, and so there is no mathematical inconsistency. 

\paragraph{Bulk Schwarzian quantum fluctuations:}

Resolving the first issue is one of the key results of this work. To do so, we make use of recent progress in computing the effect of bulk quantum fluctuations in the near-extremal regime of black holes with an emergent AdS$_2\times \mathbb{M}^2$ geometry \cite{iliesiu_statistical_2021}, where $ \mathbb{M}^2$ is a compact space, typically a sphere. These results themselves follow from the earlier discovery that near AdS$_2$ spacetimes are endowed with a Diff$(\mathbb S^1)$ invariance under time reparameterizations which is pseudo-spontaneously broken down to SL$(2,\mathbb R)$ by the AdS$_2$ state and its irrelevant deformation \cite{kitaev2015simple1,kitaev2015simple2,Maldacena:2016hyu,Maldacena:2016upp,Jensen:2016pah,Engelsoy:2016xyb}. The near-AdS$_2$ spacetimes are deformed by a universal deformation similar to that which connects the near-extremal, near-horizon AdS$_2\times \mathbb{M}^2$ geometries of black holes to their asymptotic region. Geometrically, the emergent symmetry corresponds to reparameterizations of the boundary time $\tau\mapsto u(\tau)$ in the asymptotic AdS$_2$ region. This sector of the dynamics is described by a simple effective theory given by the (Euclidean) Schwarzian action 
\begin{equation}
\label{Schwaction}
	\action{eff}[u] = - C \int_0^\beta \dd \tau_\text{E} \left\{
		u(\tau_\text{E}), \tau_\text{E}
	\right\},
	\quad
	\{u(\tau_\text{E}), \tau_\text{E}\} \equiv
	\frac{u'''}{u'} 
	- \frac{3}{2}\left( \frac{u''}{u'} \right)^2\,.
\end{equation}
Due to its simplicity, this action can be quantized and the partition function explicitly computed, \cite{Stanford:2017thb}. 

The Kaluza-Klein reduction to two spacetime dimensions is a necessary and delicate step in order to apply these results to higher-dimensional black holes. Doing so demonstrates how deviations from extremality are captured by \eqref{Schwaction}, \cite{Davison:2016ngz,Nayak:2018qej,Moitra:2018jqs,Sachdev:2019bjn,Moitra:2019bub}. Leveraging the computation of the one-loop exact Schwarzian partition function subsequently led to an explicit calculation of logarithmic contributions to the low-temperature entropy $S$ of near-extremal Reisner-Nordstr\"om black holes \cite{iliesiu_statistical_2021,Heydeman:2020hhw,Iliesiu:2022onk,Banerjee:2023quv}. In the canonical ensemble and for a compact, toroidal horizon,\footnote{The original calculation  \cite{iliesiu_statistical_2021} is presented for a spherical horizon, extending it to a compact toroidal horizon presents no difficulty.} this is 
\begin{equation}
    \vev{S}
	=
	S_0 + 4\pi^2 C T + \frac{3}{2}\log{\left(CT\right)} - \half\log{\left(\frac{2\pi}{\e^3}\right)}\,,
    \label{eq:s_sl2}
\end{equation}
where $\e$ is the Napier's constant and
\begin{equation}
	S_0
	=
	\frac{\volt}{4\ell_\text{P}^2}\frac{L^2}{r_\e^2}\,,
	\quad
	C
	=
	\frac{\volt}{4\pi\ell_\text{P}^2}\frac{L_2^2}{r_\e}\,.
    \label{eq:S0_C}
\end{equation}
$\nwtn{4}=\ell_\text{P}^2$ are Newton's constant and the Planck length, respectively.
$L$ is the radius of the AdS$_4$ spacetime, while $L_2=L/\sqrt6$ is the radius of the AdS$_2$ near-horizon near-extremal geometry. The notation $C$ for the coefficient of the Schwarzian action is a reminder that it controls the leading deviation from extremality of the entropy, i.e. the heat capacity. $\volt$ is the volume of compact transverse space, which we will take to be a torus for simplicity. It is necessary for it to be compact so that the coefficient of the Schwarzian action is not infinite. For a non-compact space, the Schwarzian contribution vanishes. 

Going beyond thermodynamics, quantum-corrected correlation functions of IR operators may be evaluated in the Schwarzian-dominated regime $\vev{\mathcal{G}^{\text{E}}}$, \cite{Mertens:2017mtv,Lam:2018pvp,Mertens:2019tcm}. $\vev{\mathcal{G}^{\text{E}}}$ needs to be connected to the correlation function of the full theory, which on the bulk side involves matching solutions in the near-horizon region to the asymptotically AdS$_4$ region \cite{Faulkner:2009wj,Davison:2013bxa}. The case of probe fermion correlation functions was studied in \cite{Liu:2024gxr}.

Unless this matching procedure has been carried out, it is not clear a priori that the relation \eqref{etas} should continue to hold after including Schwarzian quantum fluctuations, and hence one cannot simply use \eqref{eq:s_sl2} to deduce that the zero temperature limit of the shear diffusivity is no longer smooth. The shear viscosity and diffusivity must be computed anew, taking into account the bulk quantum fluctuations. 

\paragraph{Summary of results:}

Our strategy for doing so proceeds in several steps. Firstly, the shear retarded Green's function may be computed from the linearized bulk perturbations of the metric by matching \cite{Faulkner:2009wj,Davison:2013bxa}. This approach splits the bulk spacetime into two regions, an outer region far from the horizon, and an inner region close to the horizon. Provided the frequency and wavenumber of the perturbations are sufficiently small compared to the extremal radius, these two regions overlap and the inner and outer regions can be matched. 

The equation in the inner region takes the form of a massless scalar field in AdS$_2$ \cite{Faulkner:2009wj,Edalati:2009bi,Edalati:2010hk,Edalati:2010pn,Davison:2013bxa}. The retarded Green's function of this IR operator can be evaluated at the boundary of the inner region. A massless scalar in AdS$_2$ can be quantized either with $\Delta=1$ (Dirichlet) or $\Delta=0$ (Neumann) boundary conditions \cite{Klebanov:1999tb}. We find that the outer region imposes mixed boundary conditions on the inner region, consistent with choosing the $\Delta=0$ quantization. While this does not change the outcome of the matching calculation, it plays an important role in order to incorporate bulk quantum fluctuations, as the quantum-averaged correlation functions computed in \cite{Mertens:2017mtv,Lam:2018pvp,Mertens:2019tcm} depend on the dimension $\Delta$ of the operator. 

Further, the result of \cite{Mertens:2017mtv,Lam:2018pvp,Mertens:2019tcm} for generic $\Delta$ can be immediately applied to $\Delta=1$ operators, but not to $\Delta=0$ operators. This is because the classical Euclidean correlator scales like $\mathcal{G}_\Delta^{\text{E}}(\tau_\text{E})\sim 1/|\tau_\text{E}|^{2\Delta}$ for $\Delta \neq0$, but behaves logarithmically for $\Delta=0$, $\mathcal{G}_{\Delta = 0}^{\text{E}}(\tau_\text{E})\sim\log|\tau_\text{E}|$. To extract $\langle \log|\tau_\text{E}|\rangle$ in the Schwarzian theory, we carefully take a $\Delta\to0$ limit of the results of \cite{Mertens:2017mtv}. Further subtle technical steps involve analytically continuing back to real time and then Fourier transforming the quantum averaged IR retarded Green's function $\vev{\mathcal{G}_{\Delta = 0}^{\text{R}}}$.  Our final result is 
\begin{equation}
	\vev{\mathcal{G}_{\Delta = 0}^\text{R}(\omega)}
	=
	\left[
        1 
        + \frac{1}{4\pi^2}\frac{1}{CT} 
        + O\!\left(\frac1{(CT)^2}\right)
    \right]
	\mathcal{G}_{\Delta = 0}^\text{R}(\omega)
    \,.
\end{equation}
valid in a large $CT$ expansion. 

Once this is done, $\vev{\mathcal{G}^{\text{R}}_{\Delta = 0}}$ can be inserted back into the matching result, and the quantum-corrected shear viscosity and shear diffusivity follow. We find that the shear retarded Green's function takes the form
\begin{equation}
    \label{GRshearquantum}
    G^{\text{R}}_{T^{xy}T^{xy}}(\omega,k)
    =
    \frac{
        \vev{\eta}\omega^2
    }{
        \ii\omega - \vev{D_\perp} k^2
    }
\end{equation}
where
\begin{equation}
    \label{etaquantum}
    \left(
        \vev{\eta},\vev{D_\perp}
    \right)
    =
    \left(\frac{s_0}{4\pi},\frac{r_\e}{12}\right)
    \left[
        1 
        + \frac{1}{4\pi^2}\frac{1}{CT} 
        + O\!\left(\frac1{(CT)^2}\right)
    \right]
    .
\end{equation}
As we explain in detail in the main text, the leading order matching calculation needed to obtain \eqref{GRshearquantum} and \eqref{etaquantum} only captures the leading, extremal behavior of thermodynamic quantities. I.e., the entropy density in \eqref{etaquantum} $s_0=S_0/\volt$ takes its extremal value. The intuition for this is clear, in the near-extremal regime, $T r_\e\ll1$ and so $S_0\gg CT$. This implies that we should not include either deviations from extremality in $s$ or the $\log CT$ quantum corrections. To do so consistently, the matching calculation needs to be expanded one order higher, \cite{Gouteraux:2025kta,Gouteraux:2026}.

As a cross-check of our results and in order to facilitate comparison with other works \cite{Cremonini:2025yqe,Kanargias:2025vul}, we have also computed $\langle\eta\rangle$ using the Kubo formula and setting $k=0$ from the start. It is not obvious a priori that the $k\to0$ and $C
\to +\infty$ should commute. However, while the details of the calculation differ, and in particular the scaling dimension of the IR operator is in this case $\Delta=1$, we eventually recover the same leading quantum correction to $\langle\eta\rangle$, \eqref{etaquantum}, in agreement with \cite{Cremonini:2025yqe,Kanargias:2025vul}.
 
At this point, it is worth to pause in order to carefully delineate the regime of validity of our calculation. To be in the near-extremal regime, we must satisfy
\begin{equation}
    \label{nearextrregime}
    T r_\e
    \,,\,
    \omega r_\e
    \,,\,
    k r_\e
    \ll
    1
    \quad\Rightarrow\quad 
    \volt
    \gg 
    r_\e^2
    \,.
\end{equation}
In the last equality, we have taken into account the fact that the torus is compact and so $k$ is quantized in integer multiples of the volume of the torus $1/\volt$. Next, we must ask when it is sensible to interpret our result \eqref{GRshearquantum} as a hydrodynamic retarded Green's function, so that the pole and its residues can be connected to the transport coefficients and the thermodynamic quantities of the hydrodynamic theory. The hydrodynamic regime is
\begin{equation}
\omega\lesssim T\,, \quad k\lesssim \sqrt{\frac{T}{r_\e}}\quad\Rightarrow\quad \frac{\volt T}{ r_\e}\gtrsim1\,.
\end{equation}
Thanks to the near-extremal condition $Tr_\e\ll1$, we see that it suffices to take the volume of the torus sufficiently large compared to the extremal radius \eqref{nearextrregime} to be in the near-extremal, hydrodynamic regime. 

We now address the question of how important the quantum corrections due to the Schwarzian are in the hydrodynamic regime. To this extent, we evaluate whether $CT$ can be $O(1)$:
\begin{equation} \frac{r_\e}{\volt}\lesssim T\lesssim\frac1C\sim
	\frac{r_\e\ell_\text{P}^2}{\volt L^2},\quad \Rightarrow\quad L^2\lesssim \ell_\text{P}^2\,.
\end{equation}
The inequality on the left-hand side is required to be in the hydrodynamic regime. Thus, strong quantum fluctuations in the hydrodynamic regime are not compatible with keeping the AdS radius much bigger than the Planck scale, or in other words, with classical gravity in the bulk. Hence we deduce that
\begin{equation}
    \boxed{
        \begin{split}
            \textrm{Near-extremal, hydrodynamic regime: }&\frac{r_\e^2}{\volt}\lesssim Tr_\e\ll1\\
            +\quad\textrm{ Bulk classical gravity: } &\ell_\text{P}\ll  L\\
            &\\
            \Rightarrow \quad\textrm{Weak quantum fluctuations: }&CT\gg1
        \end{split}
    \nonumber}
\end{equation}

On the other hand, if we relax the requirement that we are in the hydrodynamic regime, but insist on near-extremality,
\begin{equation}
k_\eq\sim\sqrt{\frac{T}{r_\e}}\lesssim k\sim\frac{1}{\sqrt{\volt}}\ll \frac1{r_\e}\,.
\end{equation}
This implies that quantum fluctuations may be strong $CT\simeq1$, since
\begin{equation}
\frac1C\simeq\frac{r_\e\ell_\text{P}^2}{\volt L^2}\lesssim T<\frac{r_\e}{\volt}\ll\frac{1}{r_\e}
\end{equation}
is compatible with sub-Planckian curvature $\ell_\text{P}\ll L$:
\begin{equation}
\boxed{
\begin{split}
\textrm{Near-extremal, non-hydrodynamic regime: }&Tr_\e\lesssim\frac{r_\e^2}{\volt}\ll1\\
 +\quad\textrm{ Bulk classical gravity: } &\ell_\text{P}\ll  L\\
 &\\
\Rightarrow\quad \textrm{Weak or strong quantum fluctuations: }&CT\gtrsim1  
\end{split}\nonumber}
\end{equation}
Put together, these two results mean that equations \eqref{GRshearquantum} and \eqref{etaquantum} hold in both the hydrodynamic and non-hydrodynamic regimes, provided quantum fluctuations are weak.

From this point, we draw several conclusions:
\begin{enumerate} 
\item In the hydrodynamic regime, we can extract the shear viscosity from \eqref{GRshearquantum} in two independent ways, either from the Kubo formula or from the diffusivity, \eqref{etaquantum}, which both return the same result for $\vev{\eta}$ quoted in \eqref{etaquantum},
Further, quantum fluctuations tend to increase the shear viscosity as the temperature is lowered, as shown in equation \eqref{etaquantum}.  To the order our computation holds, 
\begin{equation}
    \frac{\vev{\eta}}{s_0}
    =
    \frac1{4\pi}\left[
        1 
        + \frac{1}{4\pi^2}\frac{1}{CT} 
        + O\!\left(\frac1{(CT)^2}\right)
    \right]
    \gtrsim
    \frac1{4\pi}
    \,.
\end{equation} 
Finally, the $CT\ll1$ limit diverges as $1/(CT)$ to leading order, in contrast to the classical result which had a smooth $T\to0$ limit.
\item In the quantum regime $T\lesssim \omega,k\lesssim1/r_\e$, quantum fluctuations may in principle be taken to be $O(1)$. However, here we have only performed a computation valid for weak quantum fluctuations, $CT\gg1$. As the coefficient of the quadratic term no longer has a smooth $T=0$ limit, this suffices to show that the classical $T=0$ modes are lifted by quantum corrections, even though our matching calculation is not carried out to sufficiently high order in order to determine their exact fate. 
\item Whether in the hydrodynamic regime or not, we can view the coefficient $\eta$ simply as a measure of the shear stress-tensor Kubo formula \cite{Policastro:2001yc}, which is well-known to be related to the absorption cross-section of shear gravitons \cite{Klebanov:1997kc,Gubser:1997yh}. The quantum absorption cross-section was computed in \cite{Emparan:2025sao,Biggs:2025nzs}, see also \cite{Emparan:2025qqf,Betzios:2025sct}. It was found the quantum corrections tend to increase the absorption cross-section and conjectured in \cite{Emparan:2025sao} that that quantum corrections would tend to increase the value of $\eta$ away from the classical limit set by the horizon area. Our results confirm this expectation. 
\end{enumerate}

In the remainder of the paper, we lay out the various calculations underpinning our results. In section~\ref{sec:RNAdS}, we review the Reissner-Nordstr\"om black hole and its thermodynamics. In section~\ref{sec:Matching}, we review and reformulate the matching calculation of \cite{Davison:2013bxa} using decoupled gauge-invariant variables, as well as addressing the question of the dimension of the IR operator in section~\ref{sec:IRscaling}. In section \ref{sec:QuantumcorrIRG}, we evaluate the quantum-averaged IR retarded Green's function and deduce the quantum-corrected shear retarded Green's function. We close with a discussion of our results. In appendix~\ref{app:shearhydro}, we give a lightning review of shear relativistic hydrodynamics, and refer readers to \cite{Kovtun:2012rj} for a pedagogical introduction to relativistic hydrodynamics. In appendices~\ref{app:NLO} and~\ref{app:log}, we give some extra technical details on the saddle-point approximation performed in section ~\ref{sec:QuantumcorrIRG}. In appendix~\ref{app:FT}, we Fourier transform the IR retarded Green's function directly in Euclidean space.

\paragraph{Note added:} 
Our results differ from recent works \cite{PandoZayas:2025snm,Nian:2025oei}, which also find a quantum correction to $\eta$ of order $O\!\left(1/(CT)\right)$, but with a different prefactor. It is difficult to compare directly to these works, which use a different approach than ours. \cite{Cremonini:2025yqe,Kanargias:2025vul} evaluates the quantum corrections to $\eta$ through the $k=0$ Kubo formula but without computing the full momentum-dependent dispersion relation of the pole. They find a different scaling dimension for the IR operator, $\Delta=1$, which we reproduce for clarity in section \ref{sec:dirbdycond}, but which satisfyingly gives the same correction to $\eta$ as our $k\neq0$ calculation, where $\Delta=0$.

While we find the same correction to $\eta$, we differ on the statement that the KSS bound is violated, since we argue that temperature corrections away from extremality to thermodynamic quantities, such as the entropy density and the momentum static susceptibility, must require a higher-order matching calculation.\\

\section{The Reissner-Nordstr\"om planar black hole \label{sec:RNAdS}}

We consider four-dimensional Einstein-Maxwell theory, with action:
\begin{align}
\label{eq:SEM}
  \action{EM} = {}&\int \dd^4 x\, \sqrt{-g} \left[{1 \over 2 \kappa^2} \left(R + {6 \over L^2} \right) - {1 \over 4 g_F^2} F^2 \right]\, , 
\end{align}
where we parametrize Newton's constant as $16\pi\nwtn{4}=2\kappa^2$, the cosmological constant is related to the AdS radius as $-2\Lambda=6/L^2$ and $g_F$ is the Maxwell coupling. 
The corresponding equations of motion are:\footnote{Note we are
  using the trace-reversed Einstein equations}
\begin{align}
\label{EMeqns}
  0 ={}& \partial_a \left( \sqrt{-g} F^{ab}\right)\, , & R_{ab} + {3 \over L^2} g_{ab} ={}& {\kappa^2 \over g_F^2} \left(F_a{}^c F_{bc} - {1 \over 4} g_{ab} F^2 \right)\, .
\end{align}
The lowercase Latin indices $a,b=1\dots d+2$ run over all bulk spacetime coordinates. 
The AdS Reissner-Nordstr\"om solution is given by:
\begin{align}
\label{RNsol}
  \dd \bar s^2 ={}& {L^2 \over r^2} \left( {-} f(r) \dd t^2 + {\dd r^2 \over f(r)} + \dd x^2 + \dd y^2 \right)\, , & \bar A ={}& \mu \left(1 - {r \over r_+} \right) \dd t\, , 
\end{align}
where the emblackening factor is:
\begin{align}
  f(r) = 1 - M \left(\frac{r}{r_+}\right)^3 + Q^2 \left(\frac{r}{r_+}\right)^4 \, . 
\end{align}
Here $r_+$ is the (outer) horizon radius, and $M = 1 + Q^2$ with\footnote{In the conventions of notation \cite{Edalati:2009bi,Davison:2013bxa}, ${2 g_F^2 L^2 \over \kappa^2}=1$ and $r_\text{there}=L^2/r_\text{here}$.}
\begin{align}
  Q^2 ={}& {\kappa^2 r_+^2 \mu^2 \over 2g_F^2 L^2}\,.
\end{align}
The horizon radius is set by the temperature according to:
\begin{align}
  4\pi T ={}& {-} f'(r_+) = {3 - Q^2 \over r_+} = {1 \over r_+} \left( 3 - {\kappa^2 r_+^2 \mu^2 \over 2 g_F^2 L^2 } \right)\, .
\end{align}
At extremality, we have $Q^2 = 3$,
i.e.~$r_\erm \equiv r_+(T \to 0) = {\sqrt{6}g_FL  \over \kappa\mu}$, and the
emblackening factor reads
\begin{align}
  f_\e(r) = 1 - 4 \left(\frac{r}{r_\e}\right)^3 + 3 \left(\frac{r}{r_\e}\right)^4 \, .
\end{align}
The emblackening factor can also be written more simply in terms of $r_\e$ and $r_+$ only:
\begin{align}
  f(r) = 1 - \left(\frac{r}{r_+}\right)^3 -  \frac{3r^3}{r_+r_\e^2}\left(1-\frac{r}{r_+}\right) \, .
\end{align}

This suggests that we can zoom on the near-extremal, near-horizon geometry by using the following change of coordinates
\begin{equation}
\label{changetoIRcoords}
    r_+
    =
    r_\erm-\epsilon\frac{r_\erm^2}{6\zeta_\h}\,,
    \quad 
    r
    =
    r_\erm-\epsilon\frac{r_\erm^2}{6\zeta}\,,
    \quad 
    t
    =
    \frac{\tau}{\epsilon}\,.
\end{equation}
Expanding order by order in small $\epsilon$, the solution becomes AdS$_2\times\mathbb{T}^2$:
\begin{align}
\label{AdS2T2metric}
    \dd \bar s^2 
    ={}&
    -\frac{L_2^2}{\zeta^2}\left(1-\frac{\zeta}{\zeta_\h}\right)\dd\tau^2
    +\frac{L_2^2\dd\zeta^2}{\zeta^2\left(1-\frac{\zeta}{\zeta_\h}\right)}
    +\frac{L^2}{r_\erm^2}\left(\dd x^2+\dd y^2\right) 
    + O(\epsilon)\,,\\
    \bar A_\tau(\zeta)
    ={}&
    \frac{g_FL_2}{\kappa\zeta}\left(1-\frac\zeta{\zeta_\h}\right) + O(\epsilon)\,.
\end{align}
Higher-order deformations away from  AdS$_2\times\mathbb{T}^2$ can be worked out systematically order by order in $\epsilon$, although we will not need them here.

In these coordinates, the temperature becomes
\begin{equation}
    T = \frac{\epsilon}{4\pi\zeta_\h} + O(\epsilon^2)\,.
\end{equation}

The expectation value of the density for the conserved U$(1)$ current
on the boundary is given by:
\begin{align}
  \rho \equiv{}& \langle J^t \rangle = {1 \over g_F^2} \lim_{r\to 0} \sqrt{-g} F^{rt} = {\mu \over g_F^2 r_+}  = {\sqrt2 Q L \over  \kappa g_F r_+^2}\, .
\end{align}
At extremality this becomes:
\begin{align}
  \rho_0\equiv\rho(T =0) ={}& {\mu \over g_F^2 r_\e}=\frac{\sqrt 6L}{\kappa g_F r_\erm^2}\,. 
\end{align}
The entropy density is
\begin{align}
\label{sRN}
s={}&\frac{2\pi L^2}{\kappa^2r_+^2}\quad\underset{T=0}{\Rightarrow}\quad s_0=\frac{2\pi L^2}{\kappa^2r_\erm^2}\,.
\end{align}

\section{Holographic calculation of transverse hydrodynamics \label{sec:Matching}}

We consider linear perturbations of the metric and gauge field $(g_{ab}=\bar g_{ab}+\delta g_{ab},A_a=\bar A_a+\delta A_a)$. The shear sector of the dynamics only includes the subset $\delta g_{yt}$, $\delta g_{xy}$, $\delta A_y$. Gauge redundancy under diffeomorphisms of the metric can be used to set $\delta g_{ry}=0$.
Due to invariance under spacetime translations, the perturbations can be expressed as plane waves in Fourier space
\begin{equation}
    \delta g_{ab}(r,t,x,y)
    =
    \delta g_{ab}(r)\e^{-\ii\omega t+ \ii kx}\,,
    \quad 
    \delta A_{b}(r,t,x,y)
    =
    \delta a_{b}(r)\e^{-\ii\omega t+ \ii kx}\,.
\end{equation}

Rearranging the shear fluctuations in the metric ($\delta g_{ty}$ and
$\delta g_{xy}$) and gauge field ($\delta a_y$) sectors leads to two
decoupled, gauge-invariant modes $\Phi_\pm$. At finite temperature,
this decoupling takes the form, \cite{Edalati:2010hk}:
\begin{align}
  \Phi_\pm(r) ={}& {r_+^2 \over r} {k f \over \omega^2 - k^2 f} \left[k \partial_r \delta g^y{}_t+ \omega \partial_r \delta g^y{}_x\right] 
                   - {2 Q^2 \over \mu}\left[ {2 f k^2 \over \omega^2 - f k^2} {r \over r_+} + \beta_\pm \right] \delta a_y\, , \label{eq:phi-def} \\
  \beta_\pm ={}& {3 \over 4} \left(1 + \frac{r_\e^2}{3r_+^2} \right) \left[1 \pm \sqrt{1 +{16 k^2r_\e^2 \over 27} \left(1 + {r_\e^2 \over 3r_+^2} \right)^{-2}} \right]\, . \label{eq:betapm-def}
\end{align}
Here $\delta g^y_{~t}=\bar g^{yy}\delta g_{ty}$ and $\delta g^y_{~x}=\bar g^{yy}\delta g_{xy}$.
These master fields then solve the decoupled equations:
\begin{align}
  0 ={}& f\partial_r \left(f \partial_r \Phi_\pm \right) + \left[ \omega^2 - f k^2 - {f \partial_r f \over r} - \frac{6 r\beta_\pm}{r_\e^2 r_+} f\right] \Phi_\pm\, . \label{eq:phi-eq}
\end{align}

The shear diffusion mode is captured by the $\Phi_-$ field \cite{Edalati:2010hk}, so we will focus on solving for the QNMs in $\Phi_-$. We do so by a matching procedure between an inner region defined by
\begin{equation}
r_\e-r\ll r_\e \,,
\end{equation}
which captures the dynamics close to the near-extremal horizon; and
 an outer region where
\begin{equation}
\omega^2\ll \frac{f(r)f'(r)}{r}\,,\qquad k^2\ll \frac{f'(r)}{r}\,.
\end{equation}
Since we wish to extend this outer region so that it overlaps with the inner region, we take the $r\to r_\e$ limit, upon which this condition become
\begin{equation}
    (\omega r_\e)^{2/3}\ll \frac{r_\e-r}{r_\e}\,,
    \qquad 
    (k r_\e)^2\ll \frac{r_\e-r}{r_\e}\,.
\end{equation}
Both conditions are now compatible for low enough frequencies and wavenumbers. Putting them together, we get\footnote{We thank Sameer Murthy for a discussion on this point.}
\begin{equation}
 (\omega r_\e)^{2/3}\ll \frac{r_\e-r}{r_\e}\ll1\,,\qquad  (k r_\e)^2\ll \frac{r_\e-r}{r_\e}\ll1
\end{equation}
or changing to $\zeta$ coordinates
\begin{equation}
    \label{matchingregionconditions}
    (\omega r_\e)^{2/3}\ll \frac{\epsilon r_\e}{\zeta}\ll1\,,
    \qquad  
    (kr_\e)^2\ll \frac{\epsilon  r_\e}{\zeta}\ll1\,.
\end{equation}
This implies that as long as we satisfy the conditions
\begin{equation}
 \omega r_\e\ll1\,,\qquad  (k r_\e)^2\ll1\,,
\end{equation}
so that the first side of each inequality in \eqref{matchingregionconditions} holds,
then there exists an overlap region between the outer and the inner region, which we can take to be the asymptotic AdS$_2$ region where $\zeta\to0$, while maintaining $\epsilon r_\e/\zeta\ll1$.

We will want to impose ingoing boundary conditions on the $\Phi_-$ perturbations
\begin{equation}
\Phi_- = \erm^{\ii \omega \rho} \Phi
\end{equation}
where the tortoise coordinate $\rho$ is defined by
\begin{align}
  \rho ={}& \int_0^r {\dd r' \over f(r')} \nonumber \\
  ={}&- \frac{\zeta_\h}{\epsilon}\log\left(1- \frac{\zeta}{\zeta_\h}\right)+O(\epsilon^0) \label{eq:rho-def}
\end{align}
where we used that $f(r)\sim(r_\e-r)^2$ at small temperatures and diverges in the near-horizon region. We will not need the $O(\epsilon)$ piece in this work, although it would be needed to extend the matching to the next order in $\epsilon$.

It will be convenient to define the IR rescaled tortoise coordinate
\begin{align}
    \zeta^\star  
    ={}&- 
    \frac{\zeta_\h}{\epsilon }\log\left(1- \frac{\zeta}{\zeta_\h}\right)\,.
    \label{eq:zetas-def}
\end{align}

\subsection{Inner solution}

To obtain the inner solution, we implement the scaling
\begin{align}
\label{Innerscaling}
    \omega\mapsto\epsilon \omega\,,
    \quad 
    k\mapsto\sqrt{\epsilon}k\,,
    \quad 
    r=r_\erm-\epsilon\frac{r_\erm^2}{6\zeta} \,,\quad   r_+=r_\erm-\epsilon\frac{r_\erm^2}{6\zeta_\h}
\end{align}
and expand $\Phi_-(\zeta)=\Phi_\text{in}^{(0)}(\zeta)+O(\epsilon)$. In $\zeta^\star$ coordinates, the equations for $\Phi_\text{in}^{(0)}$ is 
\begin{align}
   0 ={}& \partial_{\zeta^\star}^2 \Phi_\text{in}^{(0)} +\omega^2\Phi_\text{in}^{(0)}\, .
\end{align}
This admits a simple solution
\begin{align}
    \Phi_\text{in}^{(0)} ={}& c_- e^{-\ii \omega\zeta^\star}+c_0 e^{\ii \omega\zeta^\star}\, .
\end{align}
Imposing ingoing boundary conditions amounts to setting $c_-=0$.

We can now revert to $\zeta$ coordinates and expand in the matching region $\zeta\to0$
\begin{align}
    \Phi_\text{in}^{(0)}(\zeta)
    ={}&
    c_0\left(1-\frac{\zeta}{\zeta_\h}\right)^{-\ii\omega\zeta_\h}\\
    \underset{\zeta\to0}{=}{}&
    c_0\left(1+\ii\omega\zeta\right)+ O(\zeta^2)\,.
    \label{Innersolouterexp}
\end{align}
Observe that at this order in $\epsilon$ and $\zeta$, the effects of temperature do not play any role in the matching region.

\subsection{Outer solutions}

In the outer region, we scale 
\begin{align}
\label{Outerscaling}
    \omega\mapsto\epsilon \omega\,,
    \quad 
    k\mapsto\sqrt{\epsilon}k\,,
    \quad 
    r_+=r_\erm-\epsilon\frac{r_\erm^2}{6\zeta_\h} 
\end{align}
like in the inner region, but we do not change from $r$ to $\zeta$ coordinates. We also write
\begin{align} 
    \Phi_\text{out}
    ={}&
    \Phi_\text{out}^{(0)}
    +\epsilon \Phi_\text{out}^{(1)}
    +O(\epsilon^2)
    \,.
\end{align}
We will not need to go to higher order in $\epsilon$.
 
The equations are of the form
\begin{align}
    \frac{1}{rf_\e}\left[
        r^2f_\e\left(\frac{\Phi_\text{out}^{(n)}}r\right)'
    \right]'
    ={}&
    s_\text{out}^{(n)}[
        \Phi_\text{out}^{(0)},\dots,\Phi_\text{out}^{(n-1)}
    ]\,.
\end{align}
The sources are
\begin{align}
    s_\text{out}^{(0)}
    ={}&
    0
    \\
    s_\text{out}^{(1)}
    ={}&
    {-}k^2\frac{r_\e^3 f_\e'}{12 r^2 f_\e}
    \Phi_\text{out}^{(0)}
    + \frac{r(r-r_\e)^2}{r_\e^6\zeta_\h f_\e^2}
    (r^2+2 r r_\e+3 r_\e^2)\left(
        r\Phi_\text{out}^{(0)}{}'-\Phi_\text{out}^{(0)}
    \right)
    \,.
\end{align}

There are two linearly independent solutions to the homogeneous equation, which we take to be
\begin{align}
    \Phi_\text{out,h}^{(n)}(r)
    ={}&
    c_{n}^{R}\frac{r}{r_\e}
    + c_{n}^I\,r\int^r\frac{\dd r}{r^2f_\e}
    \,.
\end{align}
The first solution is regular at the horizon, while the second is not. Further, the integral in the second solution diverges like $1/r$ near the AdS$_4$ boundary. However, the extra factor of $r$ multiplying the integral regulates this divergence in the full solution. The integral is, explicitly:
\begin{align}
I_s\equiv\int^r\frac{\dd r}{r^2f_\e}={}&\frac{1}{6(r_\e-r)}-\frac1r+\frac5{18r_\e}\log\left(\frac{3r^2+2r r_\e+r_\e^2}{(r_\e-r)^2}\right)-\frac{23}{18\sqrt2\,r_\e}\arccot\frac{\sqrt2 r_\e}{3r+r_\e}\,.
\end{align}

After obtaining the solutions, we change from $r$ to $\zeta$ coordinates and take the small $\epsilon$ limit in order to extend the outer solutions into the overlap region with the inner solutions. Before turning to the outer solution at $O(\epsilon)$, it will pay to examine the $\epsilon\to0$ limit of the zeroth order outer solution.

Expanding $\Phi^{(0)}_\text{out}(r)$ into the matching region, we find
\begin{align}
    \Phi_\text{out}^{(0)}
    ={}&
    \frac{c_{0}^I\zeta}{r_\e\epsilon}+O(\epsilon^0)
    \,.
\end{align}
This immediately enforces $c_{0}^I=0$, since the inner solution $\Phi^{(0)}_\text{in}(\zeta)$ does not contains terms of order lower than $O(\epsilon^0)$. Hence,
\begin{align}
    \Phi_\text{out}^{(0)}(r)={}&c_{0}^{R}\frac{r}{r_\e}
\end{align}
and
\begin{align}
    s_\text{out}^{(1)}
    ={}&
    -c_0^R \frac{k^2 r_\e^2}{12r}\frac{f_\e'}{f_\e}
    \,.
\end{align}
Now, to $O(\epsilon)$, temperature effects do not occur (they do at $O(\epsilon^2)$). Hence to this order, the outer solutions are effectively at zero temperature, the effects of which are only communicated through the matching with the inner solution.
This considerably simplifies the $O(\epsilon)$ outer solution, to which we now turn. 

The Wronskian evaluates to $W=1/(r_\e f_\e)$, so that the particular solution at orders $n\geq1$ is
\begin{align}
    \Phi_\text{out,p}^{(n)}
    ={}&
    - r\int_{r_\e}^r r^2 s_\text{out}^{(n)} f_\e I_s\dd r
    + rI_s\int_{r_\e}^r \,rf_\e s_\text{out}^{(n)}\,\dd r
    \,.
\end{align}
Integrating by parts, we eventually obtain:
\begin{align}
    \Phi_\text{out}^{(1)}
    ={}&
    c_{1}^{R}\frac{r}{r_\e}
    + c_{1}^I\,r I_s
    + \frac{c_0^R}{12}(k r_\e)^2\left(1-\frac{r}{r_\e}\right)
    .
\end{align}
 
Matching the outer solution to the inner solution to order $O(\epsilon)$ gives
\begin{align}
    \label{matchingonshell}
    c_{0}^R
    ={}&
    c_0
    \,,\qquad 
    c_{1}^I
    =
    \ii \omega r_\e c_0
\end{align}
while $c_1^R$ remains undetermined at $O(\epsilon^0)$. We will not need $c_1^R$ in the context of this work, but matching at $O(\epsilon)$ will be performed in \cite{Gouteraux:2026}.

\subsection{Shear retarded Green's function \label{sec:shearclholo}}
With the matching coefficients in hand, we can now evaluate the near
(AdS$_4$) boundary for $\Phi_-$, setting $\epsilon=1$:
\begin{align}
  \Phi_-(r\to 0) ={}& {-}c_1^I+ c_0^R\frac{(k r_\e)^2}{12}+c_0^R \frac{r}{r_\e}\,.
\end{align}
This leads to the Green's function:
\begin{align}
\label{eq:Gm}
  G_-^{\text{R}} ={}&\frac{\Phi_-^\text{out}{}'(0)}{\Phi_-^\text{out}(0)}=\frac{\frac{c_{0}^R}{r_\e}}{c_{0}^R\frac{(k r_\e)^2}{12}-c_{1}^I}\,.
\end{align}
This still needs to be related to the shear retarded Green's function, which we review next.

From \eqref{eq:phi-def}, the near-boundary expansion of $\Phi_\pm$ is
given by:
\begin{align}
  \Phi_{\pm}(r\to 0) ={}& \Phi_{\pm}^{(0)} + r \Phi_\pm^{(1)} + \dotsb\, ,\\
  \Phi_\pm^{(0)} ={}& k r_+^2 \left(\omega h^{(0)}_{xy} + k h_{ty}^{(0)} \right) - {2 Q^2 \beta_\pm \over \mu} a_y^{(0)} \, , \label{eq:Phi0pm}\\
  \Phi_\pm^{(1)} ={}& 3 {k \over \omega} r_+^2 h_{xy}^{(3)} - {2Q^2 \beta_\pm \over \mu} a_y^{(1)}\, , \label{eq:Phi1pm}
\end{align}
where the $h_{xy}^{(n)}$, $h_{ty}^{(n)}$, and $a_y^{(n)}$ are the
coefficients of the fluctuations near the boundary:
\begin{align}
  \delta g_{\mu y}(r\to 0) ={}& {L^2 \over r^2} \left(h_{\mu y}^{(0)} + r^2 h_{\mu y}^{(2)}  + r^3 h_{\mu y}^{(3)} \right)+ \dotsb\, , & \delta a_y (r\to 0) ={}& a_y^{(0)} + r a_y^{(1)} + \dotsb\, .
\end{align}
Note that we've used a constraint equation in \eqref{eq:Phi1pm} which is
why $h_{ty}^{(3)}$ does not appear there, and as usual the components
$h_{\mu y}^{(2)}$ are fixed in terms of (derivatives of)
$h_{\mu y}^{(0)}$ by the equations of motion.

We can then solve for $h_{xy}^{(3)}$ in terms of $\Phi_\pm^{(1)}$ and
find:
\begin{align}
  3 h_{xy}^{(3)} ={}& {1 \over r_+^2} {\omega \over k} {\beta_+ \Phi_-^{(1)} - \beta_- \Phi_+^{(1)} \over \beta_+ - \beta_-} \xrightarrow{k\to 0} {1 \over r_+^2} {\omega \over k} \Phi_-^{(1)}\, , 
\end{align}
(recall from \eqref{eq:betapm-def} that $\beta_-(k\to 0) =O(k^2)$) and so
plugging into the (renormalized) on-shell action, we read off the
Green's function
\begin{align}
  G^{\text{R}}_{xy,xy} ={}& {-} {L^2 \over 2 \kappa^2} \omega^2 G_-^{\text{R}}\, . 
\end{align}
Combining with \eqref{eq:Gm}, we finally obtain
\begin{align}
    \label{eq:Gxyxyholo}
    G_{xy,xy}^{\text{R}} 
    ={}& 
    {-} {L^2 \over 2\kappa^2} \omega^2 G^\text{R}_- 
    = 
    {L^2 \over 2 \kappa^2 r_\erm^2} {\omega^2 \over \ii \omega - {r_\erm \over 12} k^2}\, .
\end{align}
This can be cast into the form predicted by hydrodynamics (see appendix \ref{app:shearhydro}) after the identifications
\begin{align}
\label{eq:Gxyxyholohydro}
  G_{xy,xy}^{\text{R}} 
  ={}& 
  {\eta \omega^2 \over \ii \omega - D_\perp k^2}
  \,,\quad 
  D_\perp 
  = 
  {r_\erm \over 12} 
  = 
  {\sqrt{6}g_F L  \over 12 \kappa \mu}
  \,,\quad  
  \eta 
  = 
  {L^2 \over 2 \kappa^2 r_\erm^2} 
  = 
  {\mu^2 \over 12 g_F^2 }
  \,.
\end{align}

This allows us to make an important point. We are now in position to obtain the value of the shear viscosity either from the numerator (Kubo formula) or from the denominator (location of the diffusion pole). The diffusivity is related to the shear viscosity through $D_\perp=\eta/\chi_{\pi\pi})$. 
The momentum static susceptibility $\chi_{\pi\pi}\equiv \varepsilon + p=s T+\mu\rho$ can be obtained from the Reissner-Nordstr\"om thermodynamics, see section~\ref{sec:RNAdS} and appendix~\ref{app:shearhydro}:
\begin{align}
\label{chiPiPiRN}
  \chi_{\pi\pi} ={}& sT + \mu \rho = \frac{3L^2(1+Q^2)}{2\kappa^2 r_+^3} \xrightarrow{T\to 0} \mu \rho_0 = \frac{6L^2}{\kappa^2 r_\e^3} = {2 \over \sqrt{3}} {\kappa\mu^3  \over \sqrt6 L g_F^3} \,,
\end{align}
while the shear diffusivity is \cite{Davison:2013bxa}
\begin{equation}
    \label{DetaRNT}
    D_\perp 
    = 
    \frac{r_+}{3(1+Q^2)}
    = \frac{r_+}{3\left(1+\frac{3r_+^2}{r_\e^2}\right)}
    \underset{T=0}{=}
    \frac{r_\e}{12}
    \,.
\end{equation}
Thus we see that our matching calculation, at the order in $\epsilon$ to which we are working (which is $\epsilon^0$ in the inner solution), is only sensitive to the leading (in fact, $T=0$) dependence of the diffusivity and of the static susceptibility. At this order, we can only resolve that there is a quadratically dispersing mode in the spectrum, for all frequencies and wavenumbers $\omega r_\e,k r_\e,\ll 1$. In particular, this is true whether we are in the hydrodynamic regime $\omega\lesssim T$, $k\lesssim \sqrt{T/r_\e}$, or outside of it $T\lesssim\omega\ll 1/r_\e$, $\sqrt{T/r_\e}\lesssim k\ll 1/r_\e$. This is indeed what may be observed numerically, \cite{Arean:2020eus}.

This does not imply that we did a $T=0$ calculation, but that temperature corrections only appear by going to the next order in $\epsilon$.\footnote{This point was made in \cite{Gouteraux:2025kta} for the case of the current-current retarded Green's function at zero density, and we will address it for the shear sector in \cite{Gouteraux:2026}.} At that order, we would be able to resolve that the spectrum for $\omega,k,T\lesssim1/r_\e$ is in fact more complicated, and combines a quadratically dispersing mode with a tower of gapped modes coming from the emergent SL$(2,\mathbb R)$ symmetry of the near-extremal, near-horizon AdS$_2\times\mathbb T^2$ geometry, \cite{Arean:2020eus}.

Had we kept any temperature corrections in the thermodynamic quantities appearing in \eqref{eq:Gxyxyholohydro}, then we would have obtained inconsistent expressions for the shear viscosity $\eta$, depending on whether we used the Kubo formula or the diffusivity, \eqref{etadefs}.

\section{The scaling dimension of the IR operator \label{sec:IRscaling}}

In this section, we explain how the boundary conditions inherited from the Dirichlet conditions at the AdS$_4$ boundary differ depending on whether the gravitational perturbations are solved keeping $k\neq0$ or setting $k=0$. In brief:
\begin{itemize}
    \item At $k\neq0$, we can compute the non-local shear Green's function $G^\text{R}_{xy,xy}(\omega,k)$ and extract the shear viscosity both from the location of the hydrodynamic diffusive pole, and from the $\omega\to0$, $k\\to0$ limits of the Green's function. In this case, we find that the UV Dirichlet boundary conditions lead to mixed boundary conditions in the IR, so that the IR Green's function must be interpreted as a $\Delta=0$ operator.
    \item At $k=0$, we find instead that the UV Dirichlet boundary conditions lead to IR Dirichlet boundary conditions, so that the IR Green's function is that of a $\Delta=1$ operator.
\end{itemize}
After computing the quantum-averaged IR Green's function of both $\Delta=0$ and $\Delta=1$ operators, we will ultimately show that both approaches lead to the same result, even though the order of operations $k=0$ and quantum-averaging is different, which is satisfying. 

\subsection{Mixed infrared boundary conditions for the non-local Green's function \label{sec:mixdbdycond}}

The scaling dimension of a scalar operator in AdS$_{d+2}$ is \cite{McGreevy:2009xe}
\begin{equation}
\Delta_\pm = \frac12\left(d+1\pm\sqrt{(d+1)^2+4m^2 L_{d+2}^2}\right)\,.
\end{equation}
The Breitenlohner-Friedmann bound reads in general
\begin{equation}
m^2L_{d+2}^2\geq-\frac{(d+1)^2}{4} \quad \underset{d=0}{\Rightarrow}\quad m^2L^2_2\geq-1/4
\end{equation}
for AdS$_2$. In the usual quantization scheme with Dirichlet boundary conditions, $\Delta_+\geq (d+1)/2$ or $\Delta_+\geq 1/2$ for AdS$_2$. On the other hand, the unitarity bound for AdS$_{d+2}$ is $\Delta\geq(d-1)/2$, which becomes $\Delta\geq-1/2$ for $d=0$. Some operators are missing, which can only come from the $\Delta_-$ choice, since in general $\Delta_-\leq (d+1)/2$. In AdS$_2$, the Breitenlohner-Friedmann bound imposes $\left.\Delta_-\right|_{d=0}\geq-1/2$, which is compatible with both quantization schemes.

As was remarked in \cite{Edalati:2009bi}, the leading order inner equation for the field $\Phi_-$ is that of a massless scalar in AdS$_2$, and traditionally identified as dual to a $\Delta=1$ operator in the AdS$_2$ theory, as manifest from the expansion \eqref{Innersolouterexp} near the AdS$_2$ boundary. This implies that the inner Green's function of this operator is $\mathcal{G}_{\Delta = 1}^\text{R}(\omega) = \ii\omega$.

Instead, we now show that matching the inner and outer solutions for the $\Phi_-$ perturbations imposes mixed boundary conditions on the inner region and leads to a $\Delta=0$ operator. 
Prior to imposing ingoing boundary conditions, the asymptotic AdS$_2$ solution is
\begin{equation}
    \Phi^{(0)}_\text{in}(\zeta\to0)
    =
    c_0 + c_1\zeta + O(\zeta^2)
\end{equation}
where $c_0$ and $c_1$ are the two integration constants to be interpreted as source or response. Imposing ingoing boundary conditions fixes $c_1=\ii\omega c_0$ as in \eqref{Innersolouterexp}. Reformulating matching as in \eqref{matchingonshell} without imposing the ingoing boundary condition leads to
\begin{align}
\label{matchingoffshell}
    c_{0}^R
    ={}&
    c_0
    \,,\qquad 
    c_{1}^I
    = r_\e c_1
    \,.
\end{align}
From the discussion at the beginning of section \ref{sec:shearclholo}, these constants in turn determine the near-AdS$_4$ expansion of the master field
\begin{align}
    \Phi_-^{(0)}=&\frac{c_0}{12}(kr_\e)^2- r_\e c_1\,,\quad \Phi_-^{(1)}=\frac{c_0}{r_\e}\label{IRbdyconditions}
\end{align}
where the usual Dirichlet boundary conditions for the metric and gauge field at the UV AdS$_4$ boundary instruct us that $\Phi_-^{(0)}$ is the source while $\Phi_-^{(1)}$ is the response. Hence $c_0$ is the IR response while the combination of $c_0$ and $c_1$ in \eqref{IRbdyconditions} is the IR source. This means that mixed boundary conditions apply in the IR and that the corresponding operator has $\Delta=0$ and a retarded Green's function given
\begin{equation}
        \mathcal{G}_{\Delta = 0}^{\text{R}}(\omega) =\frac1{\ii\omega}\,.
        \label{eq:holoGrDelta=0}
\end{equation}
This leads to the identification
\begin{align}
    \label{eq:GxyxyholoGIR2}
    G_{xy,xy}^{\text{R}} 
    ={}& 
    {L^2 \over 2 \kappa^2 r_\erm^2} {\omega^2 \over \frac{1}{ \mathcal{G}_{\Delta = 0}^{\text{R}}(\omega)} - {r_\erm \over 12} k^2}\, 
\end{align}
and to the Kubo formula \eqref{etadefs}
\begin{equation}
\label{KuboHolo}
    \eta 
    = 
    -\underset{\omega\to0}{\mathrm{lim}}\,\frac1\omega
    \underset{k\to0}{\mathrm{lim}}\,
    \mathfrak{Im}\,G^{\text{R}}_{xy,xy}(\omega,k)
    = 
    -{L^2 \over 2 \kappa^2 r_\erm^2} 
    \underset{\omega\to0}{\mathrm{lim}}\,
    {\omega\, \mathfrak{Im}\,
    \mathcal{G}_{\Delta = 0}^{\text{R}}(\omega)}
    =
    \frac{s_0}{4\pi}
    \,,
\end{equation}
where in the last equality we used the semi-classical value of the IR retarded Green's function. This recovers the famous result that the ratio of shear viscosity and entropy density is $1/4\pi$.

To compute quantum corrections to the shear viscosity from the non-local Green's function $G^{\text{R}}_{xy,xy}(\omega,k)$, we must evaluate 
\begin{equation}
\label{InnerIRG}
    \vev{\mathcal{G}_{\Delta = 0}^{\text{R}}(\omega)} =\vev{\frac1{\ii\omega}} 
\end{equation}
which we turn to in Section~\ref{sec:QuantumcorrIRG}.

\subsection{Dirichlet infrared boundary conditions for the shear Kubo formula \label{sec:dirbdycond}}

In other works \cite{Cremonini:2025yqe,Kanargias:2025vul}, the Kubo formula has been used to compute the shear viscosity, working directly at $k=0$ for the shear gravitational perturbation. Here we review this calculation and show that this selects $\Delta=1$ for the dimension of the IR operator. 

The shear mode of the metric, $\delta g_{xy} = \frac{L^2}{r^2}\delta g^x_{~y}(r,t)$ decouples at $k=0$ from other transverse fluctuations of the metric and gauge field. From \eqref{EMeqns}, it is straightforward to see that it obeys the equation of a massless scalar
\begin{align}
    0 
    ={}&
    \bar\nabla_a\bar\nabla^a\delta g^x_{~y} 
    \,.
\end{align}
As before, we decompose the perturbation in plane waves $\delta g^x_{~y}(r,t)=h(r)e^{-\ii\omega t}$.

\paragraph{Inner region:}\strut\\
To order $\epsilon^0$, the inner perturbation $h_\text{in}^{(0)}$ obeys the equation
\begin{align}
    \label{eq:k=0shearinner}
    0={}&
    \partial_\zeta\left[
        \left(1-\frac{\zeta}{\zeta_\h}\right)
        \partial_\zeta h_\text{in}^{(0)}
    \right]
    + \frac{\omega^2}{1-\frac\zeta{\zeta_\h}} h_\text{in}^{(0)}
    \,.
\end{align}
For the purposes of determining the boundary conditions to be applied in order to compute the inner retarded Green's function, it is useful not to impose the ingoing boundary condition on the inner solution and consider the generic asymptotics near the boundary of AdS$_2\times\mathbb T^2$, which are
\begin{align}
\label{innersolshear0UV}
    h_\text{in}^{(0)}
    \underset{\zeta\to0}{\rightarrow}{}&
    h_0 + h_1 \zeta + O(\zeta^2)
\end{align}
compatible with those of a massless scalar. As we have noted before, both quantization schemes $\Delta=0$ or $\Delta=1$ are allowed since both modes are normalizable. 

\eqref{eq:k=0shearinner} can be solved exactly by the ingoing solution
\begin{align}
    \label{innersolshear0}
    h_\text{in}^{(0)}
    ={}&
    h_0\left(1-\frac{\zeta}{\zeta_\h}\right)^{-\ii\omega \zeta_\h}
    \underset{\zeta\to0}{\rightarrow}
    h_0
    + h_0 \ii\omega\zeta
    + O(\zeta^2)
\end{align}
allowing to identify, on-shell,
\begin{align}
    & h_1=\ii\omega h_0\,.
\end{align}

\paragraph{Outer region:}\strut\\
On the other hand, the outer solution to order $O(\epsilon^2)$ obeys the following equation
\begin{align}
    0={}&
    \partial_r\left[
        r^{-2}f(r)\partial_r h_\text{out}^{(0)}
    \right]
    \,,
    \\
    0={}&
    \partial_r\left[
        r^{-2}f(r)\partial_r h_\text{out}^{(1)}
    \right]
    - \frac{r_\e r^2}{6\zeta_\h}\partial_r\left[
        r^{-1}f'(r)\partial_r h_\text{out}^{(0)}
    \right]
    \,.
\end{align}
The order $O(\epsilon^0)$ solution is 
\begin{align}
    h_\text{out}^{(0)}
    ={}&
    h_\text{out}^{(0),R}
    +h_\text{out}^{(0),I}\int_{0}^r\frac{r^2}{f(r)}\dd r
    \,.
\end{align}
The integral can be evaluated exactly and expanded in the matching region using the change from $r$ to $\zeta$ coordinates \eqref{changetoIRcoords}
\begin{align}
    \int_{0}^r\frac{r^2}{f(r)}\dd r
    ={}&
    \frac{r_\e^2\zeta}{\epsilon}
    + O(\epsilon^0)
\end{align}
from which we deduce that $h_{o}^{(0),I}=0$ when matched onto the inner solution \eqref{innersolshear0}. 
Hence the outer solution also solves the homogeneous equation at $O(\epsilon)$
\begin{align}
    h_\text{out}^{(1)}
    ={}&
    h_\text{out}^{(1),R}
    +h_\text{out}^{(1),I}\int_{0}^r\frac{r^2}{f(r)}\dd r
\end{align}
and the full outer solution is
\begin{align}
    h_\text{out}
    ={}&
    h_\text{out}^{(0),R}
    + \epsilon h_\text{out}^{(1),R}
    + \epsilon h_\text{out}^{(1),I}
    \int_{0}^r\frac{r^2}{f(r)}\dd r
    \,.
\end{align}
Matching onto the inner solution, we find 
\begin{equation}
    \label{matchinghxyk0}
    h_\text{out}^{(0),R}
    =
    h_0
    \,,\qquad 
    h_\text{out}^{(1),I}
    =
    \frac{h_1}{r_\e^2} 
    \underset{\text{on-shell}}{=}
    \frac{\ii\omega}{r_\e^2}h_0
    \,.
\end{equation}
$h_\text{out}^{(1),R}$ remains undetermined at this order in $\epsilon$. 

Equipped with the matching result \eqref{matchinghxyk0}, we can read off that the IR boundary condition remains Dirichlet, since the $r^0$ term of the outer solution matches directly onto the $\zeta^0$ term of the inner solution. Hence the scaling dimension of the IR operator is $\Delta=1$ and its retarded Green's function is
\begin{equation}
     \mathcal{G}_{\Delta = 1}^{\text{R}}(\omega)
     =
     \ii\omega
     \,.
\end{equation}
The UV retarded Green's function is
\begin{equation}
    {G}_{xy,xy}^{\text{R}}(\omega,k=0) 
    = 
    \frac{L^2}{2\kappa^2}\,
    \frac{h_\text{out}^{(1),I}}{h_\text{out}^{(0),R}}
    = 
    \frac{L^2}{2\kappa^2}\,
    \frac{\mathcal{G}_{\Delta = 1}^{\text{R}}(\omega)}{r_\e^2}
    =
    \frac{s_0}{4\pi}\ii\omega
    \,.
\end{equation}
In the last equality, we inserted the semi-classical expression for the IR retarded Green's function, which allows to recover that $\eta=s_0/(4\pi)$.

There is a crucial difference between the $k=0$ and $k\neq0$ analyses: at $k\neq0$, we found $\Delta=0$ for the IR operator, while we found $\Delta=1$ when working at $k=0$. To determine whether this has an impact on $\vev{\eta}$, we must compare $\vev{\mathcal{G}_{\Delta = 0}^{\text{R}}(\omega)}$ and $\vev{\mathcal{G}_{\Delta = 1}^{\text{R}}(\omega)}$.

\section{Computing the quantum-averaged inner Green's function \label{sec:QuantumcorrIRG}} 

We have determined that the IR operator which controls quantum corrections to the shear response can have $\Delta=0$ or $\Delta=1$, depending on whether the wavenumber is set to zero or not at the start of the calculation. We will now compute the quantum-averaged retarded Green's function for these operators using the 1-loop exact path integral over the effective Schwarzian action. This will involve several technical steps, which we explain in detail in this section. 

Before we do so, we pause to comment on the applicability of the $0+1$-dimensional Schwarzian quantum-corrected correlators to higher-dimensional black holes. As we alluded to in the introduction, this involves a Kaluza-Klein reduction along the transverse torus $\mathbb{T}^2$ down to two spacetime dimensions. In order to truncate the low-energy effective theory to the Schwarzian action \eqref{Schwaction}, the massless and massive modes of the torus must decouple. The massive modes do decouple thanks to the property that the $O(\epsilon^0)$ inner retarded Green's functions $\mathcal{G}^{\text{R}}_{\Delta=0,1}(\omega)$ does not depend on $k$ (irrespective of the choice of quantization). The contribution of the massless modes may be pushed down to lower temperatures than $1/C$ provided we take $r_\e L\lesssim1$, \cite{iliesiu_statistical_2021}, i.e. sufficiently large black holes in AdS$_4$.

\subsection{Euclidean and Retarded Two-point Functions}

Consider coupling a scalar operator $\mathcal{O}_\Delta$ with source $\Phi_0$ and scaling dimension $\Delta$ to the Schwarzian theory governing the near AdS$_2$ dynamics. 
The length of the boundary is fixed by the inverse temperature of the black hole $\beta\equiv1/T$ seen from the outer region. 
The two-point function of $\mathcal{O}_\Delta$, that we refer to also as the bilocal operator $\mathcal{G}^\text{E}_\Delta$, is naturally expressed in a frame with explicit $\beta$-periodicity and reads,%
\footnote{Keeping a multiplicative constant that sets the normalisation obtained from holography.}
\begin{equation}
    \mathcal{G}^\text{E}_\Delta(\tau^\text{E}_1,\tau^\text{E}_2)
    =
    \mathcal{N}_\Delta
    \left(
		\frac{
			u'(\tau^\text{E}_1) u'(\tau^\text{E}_2)
		}{
			\frac{\beta^2}{\pi^2}\sin^2(\frac{\pi}{\beta}|u(\tau^\text{E}_1) - u(\tau^\text{E}_2)|)		
		}	
	\right)^{\!\Delta}\,,
    \label{eq:bilocal}
\end{equation}
where $u\in \mathrm{Diff}(\mathbb{S}^1)$, $u>0$ corresponds to a time reparameterization of the thermal thermal circle.
The Schwarzian theory being SL$(2,\mathbb{R})\subset \mathrm{Diff}(\mathbb{S}^1)$ invariant, the physically distinguishable configurations are reparameterizations $u\in \mathrm{Diff}(\mathbb{S}^1)/\mathrm{SL}(2,\mathbb{R})$.
In order to compute quantum corrections to the two-point function, we define the following generating functional in the path integral formulation:
\begin{equation}
    \mathcal{Z}[\Phi_0]
    =
    \underset{\mathrm{Diff}(\mathbb{S}^1)/\mathrm{SL}(2,\mathbb{R})}{\int}[\mathcal{D}u]\,
    \e^{
        S_0 
        + C\int_0^\beta\dd\tau_\text{E}\,\left\{\tan\frac{\pi u(\tau_\text{E})}{\beta},\tau_\text{E}\right\} 
        + \half\int\dd\tau^\text{E}_1\dd\tau^\text{E}_2\,\Phi_0(\tau^\text{E}_1)\mathcal{G}^\text{E}_\Delta(\tau^\text{E}_1,\tau^\text{E}_2)\Phi_0(\tau^\text{E}_2)
    }
    \,.
\end{equation}
The constants $S_0$ and $C$ are respectively the extremal Bekenstein-Hawking entropy of the black hole and the Schwarzian coupling.
Turning off the source, the generating functional reduces to the partition function of the Schwarzian theory, which is known exactly \cite{Mertens:2022irh}:
\begin{equation}
    \left.\mathcal{Z}[\Phi_0]\right|_{\Phi_0=0}
    \equiv
    \mathcal{Z}(\beta)
    =
    \frac{1}{4\pi^2}\left(\frac{2\pi C}{\beta}\right)^{\!3/2}\e^{S_0+\frac{2\pi^2C}{\beta}}\,.
\end{equation}
The quantum average expectation value of the bilocal operator $\mathcal{G}^\text{E}_\Delta$  is extracted by taking functional derivatives with respect to the source $\Phi_0$ of the operator $\mathcal{O}_\Delta$:
\begin{equation}
    \vev{\mathcal{G}^\text{E}_\Delta(\tau^\text{E}_1,\tau^\text{E}_2)}
    =
    \frac{1}{\mathcal{Z}[\Phi_0]}\left.
        \frac{\delta^2\mathcal{Z}[\Phi_0]}{\delta\Phi_0(\tau^\text{E}_1)\delta\Phi_0(\tau^\text{E}_2)}
    \right|_{\Phi_0=0}
    \,.
    \label{eq:functional_derivatives}
\end{equation}
The general form of these corrections has already been obtained in \cite{Mertens:2017mtv} and reads:
\begin{equation}
	\vev{\mathcal{G}_\Delta^\text{E}(\tau_\text{E})}
	=
	\mathcal{N}_\Delta \frac{\e^{-S_0}}{\mathcal{Z}(\beta)}
	\int\dd \mu(E_1) \dd \mu(E_2)\,
	\e^{-|\tau_\text{E}|E_1}\e^{-(\beta - |\tau_\text{E}|)E_2}
	\frac{
		2\Gamma(\Delta \pm \ii\sqrt{2CE_1} \pm \ii\sqrt{2CE_2})
	}{
		(2C)^{2\Delta}\Gamma(2\Delta)
	}\,,
	\label{eq:vev_2pt}
\end{equation}
where $\tau_\text{E} = \tau^\text{E}_1-\tau^\text{E}_2$, 
$\dd\mu(E) = \dd E\,C\e^{S_0} \sinh(2\pi\sqrt{2CE})/(2\pi^2)$ for $E$ varying between $0$ and $+\infty$, and $\Gamma(\Delta \pm \ii\sqrt{2CE_1} \pm \ii\sqrt{2CE_2})$ is a short-hand notation for the product of the four different Gamma functions corresponding to the different signs of the $\pm$'s.
As the quantum average of the bilocal operator only depends on the Euclidean time through $\tau_\text{E}$, we take the liberty to write it with a single variable $\vev{\mathcal{G}^\text{E}_\Delta(\tau_\text{E})}$ for ease of notation.
The expression in Eq.~\eqref{eq:vev_2pt} is defined for $\lvert\tau_\text{E} \rvert\in(0,\beta)$ only, and should be understood beyond this interval by a periodic continuation $\vev{\mathcal{G}^\text{E}_\Delta(\tau_\text{E}+\beta)}=\vev{\mathcal{G}^\text{E}_\Delta(\tau_\text{E})}$.
Note also that the transformation $\tau_\text{E}\to \beta-\tau_\text{E}$ leaves the correlator invariant, reflecting the symmetry under the exchange $E_1\leftrightarrow E_2$.

\subsection{Saddle-point Approximation}

The correlator \eqref{eq:vev_2pt} can be evaluated by performing a saddle-point approximation in the large $C$ limit.
In order to find the saddle-point, we use the change of variable $(E_1,E_2)\to (M,\omega)$, inspired by \cite{Lam:2018pvp}, and defined as%
\footnote{$\omega$ effectively corresponds to the Lorentzian frequency after a Wick rotation from Euclidean to Lorentzian time, see~\ref{ssec:Fourier}.}
\begin{equation}
	E_1 
	= 
	M + \frac{\omega}{2}
	\,,
	\qquad
	E_2 
	= 
	M - \frac{\omega}{2}
	\,.
\end{equation}
We make this choice of change of variables to manifestly preserve the symmetry $\tau_\text{E}\to \beta-\tau_\text{E}$, encapsulated in the symmetry under the exchange $E_1\leftrightarrow E_2$.
In terms of the variables $(M,\omega)$, the exchange is now obtained by changing the sign of $\omega$.
Interpreting $E_1$ and $E_2$ as energies of two given states, $M$ is understood as their average and $\omega$ as their difference.
The two-point function now reads:
\begin{equation}
	\vev{ \mathcal{G}_{\Delta}^\text{E}(\tau_\text{E}) }
	=
	\mathcal{N}_\Delta	
	\frac{C^2}{\pi^2}\left(\frac{\beta}{2\pi C}\right)^{3/2}\e^{-\frac{2\pi^2 C}{\beta}}
	\int\dd \omega\,
	\e^{(\beta/2 - |\tau_\text{E}|)\omega}
	\int\dd M\, \e^{-F(M;C)}\,,
	\label{eq:two-pt_Delta}
\end{equation}
where we define a function $F$ from the integrand of \eqref{eq:vev_2pt} as:
\begin{equation}
	\e^{-F(M;C)}
	=
	\e^{-\beta M}
	\sinh{(2\pi \sqrt{C(2M\pm \omega)})}
	\frac{ 
		2\Gamma(\Delta \pm \ii \sqrt{C(2M + \omega)} \pm \ii \sqrt{C(2M - \omega)} )
	}{
		(2C)^{2\Delta}\Gamma(2\Delta)
	}\, .
	\label{eq:F_saddle}
\end{equation}
Again, $\pm$ in the argument of the hyperbolic sine function is a shorthand notation to denote the product of the two functions with different signs.
The integral over $M$ is strongly peaked around $M^\star$ and as such can be calculated using a saddle-point approximation:
\begin{equation}
	\vev{ \mathcal{G}_{\Delta}^\text{E}(\tau_\text{E}) }
	=
	\mathcal{N}_\Delta
	\frac{C^2}{\pi^2}\left(\frac{\beta}{2\pi C}\right)^{3/2}\e^{-\frac{2\pi^2 C}{\beta}}
	\int\dd \omega\,
	\e^{(\beta/2 - |\tau_\text{E}|)\omega}
	\sqrt{
		\frac{2\pi}{\partial_M^2 F(M^\star;C)}
	} \e^{-F(M^\star;C)}\,,
	\label{eq:two-pt_saddle}
\end{equation}
where the saddle-point $M^\star(\omega)$ is the point at which $F$ is minimum, and in general depends on $\omega$.

In order to evaluate \eqref{eq:two-pt_saddle}, we now look for the minimum of $F$.
From the symmetry under the exchange $E_1\leftrightarrow E_2$, we expect the saddle-point energies $E_1^\star$ and $E_2^\star$ to be the same, which motivates neglecting the difference $\omega$ with respect to the average $M$. In the regime of weak quantum fluctuations, i.e.~$CT \gg 1$, the integral over $M$ is dominated to leading order in $CT$ by a saddle-point at $M^\star=2\pi^2 C/\beta^2$, which satisfies the conditions listed in this paragraph.
This leading-order result ultimately, after performing the remaining integral over $w$ (see \eqref{eq:GEDelta_tree} below), reproduces \eqref{eq:bilocal} with $u$ being the trivial diffeomorphism.
Stated differently, the semi-classical limit of \eqref{eq:vev_2pt} is simply the thermal two-point function in a CFT$_1$.


Instructed by the leading order saddle-point approximation, we want to go beyond in order to capture quantum corrections.
To this aim, we need to extract $M^\star$ in a perturbative expansion for which $CM^\star\gg 1$, $M^\star\gg \omega$, $\beta/C \ll 1$, with $\beta\omega$ non-perturbative.
We therefore introduce the following dimensionless variables:
\begin{equation}
	m 
	= 
	\frac{\beta^2}{C}M\,,
	\qquad
	w 
	= 
	\beta\omega\,,
	\qquad
	x 
	= 
	\frac{\beta}{C}\,.
\end{equation}
$x$ should be understood as our perturbative parameter, with $x=0$ corresponding to the semi-classical result.
The perturbative expansion is carried out by expanding 
$m = m^{(0)} + x\, m^{(1)} + x^2\, m^{(2)} + \ldots$ as $x\to 0$ while keeping $w$ fixed.
The saddle-point is found,
after neglecting exponentially suppressed corrections as $x\to 0$ and defining $s_\pm(x) = \sqrt{2m \pm x w}$, by solving
\begin{equation}
	\begin{split}
	0 \overset{!}=
	\beta^{-1}\partial_M F(M;C)
	=&~
	1 - \frac{2\pi}{s_+} - \frac{2\pi}{s_-}
	+ 2\left(
		\frac{1}{s_+} + \frac{1}{s_-}
	\right)\mathfrak{Im}\,\psi\!\left[\Delta +\ii(s_+ + s_-)/x\right]
	\\
	&~	
	+ 2\left(
		\frac{1}{s_+} - \frac{1}{s_-}
	\right)\mathfrak{Im}\,\psi\!\left[\Delta +\ii(s_+ - s_-)/x\right]
	,
	\end{split}
	\label{eq:saddle-eq}
\end{equation}
Expanding \eqref{eq:saddle-eq} as $x\to 0$, the first two terms in the $x$ expansion of the saddle-point read
\begin{equation}
	m^{(0)}
	=
	2\pi^2\,,
	\qquad
	m^{(1)}
	= 2\Delta - 1
	+ \frac{w}{\pi}\mathfrak{Im}\,
	\psi\!\left[\Delta + \ii w/2\pi\right]
	.
	\label{eq:saddle_Delta}
\end{equation}
The quantum-averaged correlator \eqref{eq:saddle_Delta} can now be expressed up to next-to-leading order in $x$:
\begin{equation}
	\begin{split}
	\vev{\mathcal{G}^\text{E}_{\Delta}(\tau_\text{E})}
	=&~
	\frac{\mathcal{N}_\Delta}{2\pi \beta}\left(\frac{2\pi}{\beta}\right)^{2\Delta - 1}
	\int\dd w\,
	\e^{w(\beta/2 - |\tau_\text{E}|)/\beta}
	\frac{\Gamma(\Delta \pm \ii w/2\pi)}{\Gamma(2\Delta)}\times
	\\
	&~\left(
		1 + x\left[
			\frac{\left(m^{(1)}\right)^2}{8\pi^2}
			- \frac{w^2}{32\pi^2}
			+ \frac{2\Delta - 1}{8\pi^2}
			- \frac{w^2}{16\pi^4}\mathfrak{Re}\,\psi'\!\left[
				\Delta + \ii w/2\pi
			\right]
		\right] + O(x^2)
	\right)
	\\
	\equiv&~
	\vev{\mathcal{G}_\Delta^\text{E}(\tau_\text{E})}_\text{tree}
	+ \vev{\mathcal{G}_\Delta^\text{E}(\tau_\text{E})}_\text{one-loop}
	+ O(x^2)
	\,.
	\end{split}
	\label{eq:GEDelta_saddle}
\end{equation}
We are now left with evaluating the integral over $w$. 
This is detailed in the next subsection.

\subsection{Fourier Space Correlator}
\label{ssec:Fourier}

The integral in \eqref{eq:GEDelta_saddle} can be obtained by analytically continuing $\tau_\text{E}\to \ii \tau + 0^+$ for $\tau_\text{E} > 0$ (respectively $\tau_\text{E}\to \ii \tau - 0^+$ for $\tau_\text{E} < 0$), which maps the Euclidean correlator $\mathcal{G}_\Delta^\text{E}(\tau_\text{E}>0)$ to the real-time Wightman correlator $\mathcal{G}_\Delta^-(\tau)$ (respectively $\mathcal{G}_\Delta^\text{E}(\tau_\text{E}<0)$ to $\mathcal{G}_\Delta^+(\tau)$), as illustrated in Figure~\ref{fig:Wick_rotation}.
After Wick rotation, the remaining integral to evaluate is now an inverse Fourier transform.

\begin{figure}[htbp]
	\centering
	\begin{tikzpicture}
		\draw [help lines,->] (0,-2) -- (0,2) coordinate (yaxis);
		\node [below right] {$0$};
		\node [right] at (yaxis) {$-\ii\tau_\text{E}$};
		\draw [thick, color = myblue] (0,0) -- (0,2);
		\draw [thick, color = myblue] (0,0) -- (0,-2);
		\node at (-1.1,1.5) {{\color{myblue}$\mathcal{G}_\Delta^\text{E}(\tau_\text{E}<0)$}};
		\node at (-1.1,-1.5) {{\color{myblue}$\mathcal{G}_\Delta^\text{E}(\tau_\text{E}>0)$}};
		\draw [->] (-2.2,-1.5) arc (-90:-165:0.5);
		\node at (-3.7,-1.6) {$\tau_\text{E}\to \ii\tau + 0^+$};
		\draw [->] (-2.2,1.5) arc (90:165:0.5);
		\node at (-3.7,1.6) {$\tau_\text{E}\to \ii\tau - 0^+$};
		\draw[dashed] (3.5,0.5) -- (2.7,0.5) node[left] {$+\ii 0^+$};
		\draw[dashed] (3.5,-0.5) -- (2.7,-0.5) node[left] {$-\ii 0^+$};
		\draw [decorate, decoration={zigzag, segment length = 7, amplitude = 2}, myred]
			(-3.5, 0) -- (3.5, 0);
		\draw (0,0) node {{\color{myred}$\bullet$}};
		\node [above right] at (3.5, 0) {$\tau$};
		\draw [dashed] (-3.5,0.5) -- (1.55,0.5) 
			node [above, very near start] {$\mathcal{G}_\Delta^+(\tau)$};
		\draw [dashed] (-3.5,-0.5) -- (1.55,-0.5) 
			node [below, very near start] {$\mathcal{G}_\Delta^-(\tau)$};
	\end{tikzpicture}
	\caption{By performing a Wick rotation with Euclidean time $\tau_\text{E} > 0$ (respectively $\tau_\text{E} < 0$), one obtains respectively the Wightman correlator $\mathcal{G}_\Delta^{-}$ (respectively $\mathcal{G}_\Delta^{+}$).}
	\label{fig:Wick_rotation}
\end{figure}

\subsubsection{Semi-Classical Analysis}
\label{sssec:semi-classical}

Let us focus first on evaluating the tree level correlator, since the procedure is the same at the next order while expressions are more tractable.
The Wick rotation of \eqref{eq:GEDelta_saddle}, taking $\tau_\text{E}>0$ for concreteness, yields, working in terms of $\omega = w/\beta$:
\begin{equation}
	\vev{\mathcal{G}_\Delta^-(\tau)}_\text{tree}
	=
	\frac{\mathcal{N}_\Delta}{2\pi}
	\left(\frac{2\pi}{\beta}\right)^{2\Delta-1}
	\int_{-\infty}^{+\infty}\dd\omega\,
	\e^{-\ii\omega (\tau-\ii 0^+)}
	\e^{\beta\omega/2}
	\frac{\Gamma(\Delta\pm \ii\beta\omega/2\pi)}{\Gamma(2\Delta)}
	\,.
	\label{eq:inv_FT_Delta}
\end{equation}
The $\ii 0^+$ prescription makes sure that the integrand is exponentially suppressed as $\omega \to +\infty$. 
One recognises an inverse Fourier transform of the momentum space Wightman correlator, and as such, $\omega$ can be interpreted as the Lorentzian frequency.
To make further progress, we use the residue theorem.
The poles of the integrand are located at $\omega_n^{\pm} = \pm 2\ii\pi(\Delta + n)/\beta, n \in \mathbb{N}$.
\begin{figure}[htbp]
	\centering
	\begin{tikzpicture}
		\draw [help lines,->] (-3.5,0) -- (3.5,0) coordinate (xaxis);
		\draw [help lines,->] (0,-3.5) -- (0,3.5) coordinate (yaxis);
		\node [below left] {$0$};
		\node [above right] at (xaxis) {$\mathfrak{Re}\,\omega$};
		\node [left] at (yaxis) {$\mathfrak{Im}\,\omega$};
		\node at (0,1) {$\times$};
		\node at (0,2) {$\times$};
		\node at (0,3) {$\times$};
		\node at (.7,1) {$\frac{2i\pi}{\beta}\Delta$};
		\node at (1.2,2) {$\frac{2i\pi}{\beta}(\Delta + 1)$};
		\node at (0,-1) {$\times$};
		\node at (0,-2) {$\times$};
		\node at (0,-3) {$\times$};
		\node at (.7,-1) {$-\frac{2i\pi}{\beta}\Delta$};
		\node at (1.2,-2) {$-\frac{2i\pi}{\beta}(\Delta + 1)$};
		\draw [->, line width=0.8pt] (-3.2,0) -- (-1.6,0);
		\draw [->, line width=0.8pt] (-1.6,0) -- (1.6,0);
		\node [below left] at (-3.2, 0) {$-R$};
		\node [below right] at (3.2, 0) {$R$};
		\node [above] at (-1.6, 0) {$\mathscr{I}_R$};
		\draw [->, line width=0.8pt, color = myred] (0,0) -- (3.2,0) arc (0:45:3.2);
		\draw [->, line width=0.8pt, color = myred] ({3.2/sqrt(2)},{3.2/sqrt(2)}) arc (45:135:3.2);
		\draw [line width=0.8pt, color = myred] ({-3.2/sqrt(2)},{3.2/sqrt(2)}) arc (135:180:3.2);
		\node at (-2,3) {{\color{myred}$\mathscr{C}_R^+$}};
		\draw [->, line width=0.8pt, color = myblue] (0,0) -- (3.2,0) arc (0:-45:3.2);
		\draw [->, line width=0.8pt, color = myblue] ({3.2/sqrt(2)},{-3.2/sqrt(2)}) arc (-45:-135:3.2);
		\draw [line width=0.8pt, color = myblue] ({-3.2/sqrt(2)},{-3.2/sqrt(2)}) arc (-135:-180:3.2);
		\node at (-2,-3) {{\color{myblue}$\mathscr{C}_R^-$}};
	\end{tikzpicture}
	\caption{Different integration contours in the complex frequency plane to compute the inverse Fourier transform Eq.~\eqref{eq:inv_FT_Delta}.}
	\label{fig:inv_FT_Delta}
\end{figure}
The choice of how to close the contour depends on the sign of the Lorentzian time $\tau$.
The contours are drawn in Figure~\eqref{fig:inv_FT_Delta}: we consider $\mathscr{I}_R\cup\mathscr{C}_R^+$ for $\tau < 0$ and $\mathscr{I}_R\cup\mathscr{C}_R^-$ for $\tau > 0$.
As the contour is pushed towards infinity, the integrand along the $\mathscr{C}_R^\pm$ vanishes, reducing the contour integral to an integral along the real line, i.e. to \eqref{eq:inv_FT_Delta}.
Focusing on $\tau > 0$, one has from the residue theorem
\begin{equation}
	\vev{\mathcal{G}_\Delta^-(\tau)}_\text{tree}
	=
	-\ii\mathcal{N}_\Delta\left(\frac{2\pi}{\beta}\right)^{2\Delta - 1}
	\sum_{n\in\mathbb{N}}
	\underset{\omega = \omega^-_{n}}{\mathrm{Res}}\left(
		\e^{-\ii\omega(\tau-\ii 0^+)}
		\e^{\beta\omega/2}
		\frac{\Gamma(\Delta\pm \ii\beta\omega/2\pi)}{\Gamma(2\Delta)}
	\right)
	\,.
	\label{eq:residue_th}
\end{equation}
The overall minus sign on the right hand side comes from the clock-wise orientation of the contour.
The residues evaluate to
\begin{equation}
	\underset{\omega = \omega^-_{n}}{\mathrm{Res}}\left(
		\e^{-\ii\omega(\tau-\ii 0^+)}
		\e^{\beta\omega/2}
		\frac{\Gamma(\Delta\pm \ii\beta\omega/2\pi)}{\Gamma(2\Delta)}
	\right)
	=
	\ii\frac{2\pi}{\beta}
	\left(-\ii\e^{-\frac{\pi}{\beta}(\tau-\ii 0^+)}\right)^{2\Delta}
	\frac{(2\Delta)_n}{n!} 
	\left(\e^{-\frac{2\pi}{\beta}(\tau-\ii 0^+)}\right)^n,
\end{equation}
where we used the fact that the $(n+1)^\text{th}$ pole of the Gamma function is $(-1)^n/n!$ and introduced the Polchhammer's symbol $(a)_n=\Gamma(a+n)/\Gamma(n)$.
Plugging this result into~\eqref{eq:residue_th}, one recognises the entire series:
\begin{equation}
	\vev{\mathcal{G}_\Delta^-(\tau)}_\text{tree}
	=
	\mathcal{N}_\Delta
	\left(
		\frac{\pi}{\beta}
		\frac{1}{\ii\sinh \frac{\pi}{\beta}(\tau-\ii 0^+)}
	\right)^{2\Delta}
	\quad
	\text{for } \tau > 0\,.
	\label{eq:Wightman_tree}
\end{equation}
Doing exactly the same analysis assuming now $\tau<0$ leads to the same Wightman correlator as \eqref{eq:Wightman_tree}.
The Euclidean correlator $\mathcal{G}_\Delta^\text{E}$ is recovered after Wick rotating back to Euclidean time,
\begin{equation}
	\vev{\mathcal{G}^\text{E}_\Delta(\tau_\text{E})}_\text{tree}
	=
	\mathcal{N}_\Delta
	\left(
		\frac{\pi}{\beta\sin(\frac{\pi}{\beta}\tau_\text{E})}
	\right)^{2\Delta}
	\,,
	\label{eq:GEDelta_tree}
\end{equation}
which is, as claimed earlier, the semi-classical Euclidean two-point function.
The same analysis can be repeated assuming $\tau_\text{E}<0$ instead, and we find in that case that $\tau_\text{E}$ in the right hand side of~\eqref{eq:GEDelta_tree} is replaced by $|\tau_\text{E}|$.

\subsubsection{Beyond the Semi-Classical Analysis}

The one-loop corrections can be computed in a similar fashion, the pole being at the same locations, but with more cumbersome expressions:
now the integrand also has triple poles.
In Appendix~\ref{app:NLO}, we follow the same steps as above (namely Wick rotating, summing over residues, and Wick rotating back) and find
\begin{equation}
	\begin{split}
	\frac{
		\vev{\mathcal{G}^\text{E}_\Delta}_\text{one-loop}
	}{
		\vev{\mathcal{G}^\text{E}_\Delta}_\text{tree}
	}
	=&~
	\Delta\frac{\beta}{16\pi^2 C}
	\bigg[
		\left( 
			2(2\Delta + 1) 
			+ \Delta
			\frac{4\pi^2}{\beta^2}|\tau_\text{E}|(\beta - |\tau_\text{E}|)
		\right)
		\left( 1 - \cot(\pi|\tau_\text{E}|/\beta)^2 \right)
	\\
	&~
		+ (2\Delta + 1)\frac{8\pi}{\beta}
		\frac{
			\beta/2-|\tau_\text{E}|
		}{
			\tan{(\pi|\tau_\text{E}|/\beta)}
		}
		+ \frac{
			2(2\Delta + 1) 
			- (\Delta + 1)
			\frac{4\pi^2}{\beta^2}|\tau_\text{E}|(\beta - |\tau_\text{E}|)
		}{
			\sin(\pi|\tau_\text{E}|/\beta)^2
		}
	\bigg]
	\,.
	\end{split}
	\label{eq:Delta_loop}
\end{equation}
This result agrees with the one-loop correction (4.36) of \cite{Maldacena:2016upp} for $\beta = 2\pi$.
In the following, two specific cases are explored: $\Delta = 1$ and $\Delta = 0$.

\paragraph{$\boldsymbol{\Delta=1}$}\strut\\
Computing the Matsubara coefficients is achieved by introducing a regulator $\epsilon$,
\begin{equation}
    \vev{\mathcal{G}_\Delta^\text{E}(\omega_n^\text{E})}
    =
    \underset{\epsilon\to 0}{\mathrm{lim}}\,
    \int_\epsilon^{\beta-\epsilon}\dd\tau_\text{E}\,
    \e^{\ii\omega_n^\text{E}\tau_\text{E}}
    \vev{\mathcal{G}_\Delta^\text{E}(\tau_\text{E})}
    \,,
    \qquad
    \omega_n^\text{E}
    =
    \frac{2\pi n}{\beta}
    \,.
    \label{eq:Matsubara_coef}
\end{equation}
For $\Delta = 1$, and fixing $\mathcal{N}_\Delta = 1/\pi$ to agree with the normalization set by the holographic computation, we find up to next-to-leading order,
\begin{equation}
    \begin{split}
    \vev{\mathcal{G}_\Delta^\text{E}(\omega_n^\text{E})}
    =&~
    \frac{2}{\pi\beta}\epsilon^{-1}
    -|\omega_n^\text{E}|
    \\
    &~+ \frac{\beta}{C}\left(
		\frac{1}{6\pi\beta}
		+ \frac{|\omega_n^\text{E}|}{4\pi^2}
		+ \frac{\beta |\omega_n^\text{E}|^2}{4\pi^3}
		- \frac{\beta^2 |\omega_n^\text{E}|^3}{8\pi^4}\psi'(\beta |\omega_n^\text{E}|/2\pi)
	\right)
    \\
	&~ + O(\beta^2/C^2, \epsilon)
    \,.
    \end{split}
    \label{eq:Matsubara_coef_reg}
\end{equation}
Note that the divergent term in $\epsilon$ comes from the leading order, semi-classical result.
The retarded Green's function is then obtained through the following steps: firstly truncate \eqref{eq:Matsubara_coef_reg} by removing divergences in the regulator $\epsilon$, secondly send $\epsilon$ to zero, and finally analytically continue to the retarded Green's function by restricting to the positive Matsubara frequencies using Carlson's theorem,
\begin{equation}
    \vev{\mathcal{G}^\text{R}_{\Delta = 1}(\omega = \ii\omega_{n>0}^\text{E})}
	=
    \vev{\mathcal{G}^\text{E}_{\Delta = 1}(\omega_{n>0}^\text{E})}\,.
\end{equation}
Such a regularization goes under the name of finite part regularization and the claim is that the Euclidean correlator defined through this gives the correct retarded Green's function after analytic continuation. 
This fact is illustrated for the semi-classical result at generic $\Delta$ in Appendix~\ref{app:FT}.
We eventually obtain
\begin{equation}
    \begin{split}
	\vev{\mathcal{G}^\text{R}_{\Delta = 1}(\omega)}
	=&~
    \vev{\mathcal{G}^\text{E}_{\Delta = 1}(\omega_{n>0}^\text{E})}
    \\
	=&~\ii \omega
	+
	\frac{\beta}{C}\left[
		\frac{1}{6\pi\beta}
		- \frac{\ii \omega}{4\pi^2}
		- \frac{\beta \omega^2}{4\pi^3}
		- \ii\frac{\beta^2 \omega^3}{8\pi^4}\psi'\left(-\frac{\ii\beta \omega}{2\pi}\right)
	\right]
	+ O\left(\frac{\beta^2}{C^2}\right)
	,
    \end{split}
\end{equation}
where $\psi'$ denotes the trigamma function.
In the hydrodynamic regime, that is for $\beta\omega \ll 1$, the above expression reduces to
\begin{equation}
	\vev{\mathcal{G}^\text{R}_{\Delta = 1}(\omega)}
	\underset{\beta\omega\to 0}=
	\ii \omega
	+ \frac{1}{6\pi C}
	+ \ii\frac{\beta \omega}{4\pi^2 C}
	+ O(\beta^2/C^2,\beta^2\omega^2)
	\,.
    \label{eq:vevG1}
\end{equation}
We will comment on the presence of the real part in the IR Green's function later.

\paragraph{$\boldsymbol{\Delta=0}$}\strut\\
The case $\Delta = 0$ needs special treatment.
For this discussion, we fix $\mathcal{N}_\Delta = 1$.
Naively sending the conformal dimension to zero, \eqref{eq:vev_2pt} goes to 1.
In order to obtain the quantum-averaged correlator we are interested in, we recall that the momentum space retarded and Euclidean Green's functions of a `CFT$_1$' scalar operator with scaling dimension $\Delta = 0$ are:\footnote{The overall normalization is fixed by the holographic calculation of the correlator, \eqref{eq:holoGrDelta=0}.}
\begin{equation}
	\mathcal{G}^\text{R}_{\Delta=0}(\omega)
	=
	\frac{1}{\ii \omega}
	\,,
	\qquad
	\mathcal{G}^\text{E}_{\Delta=0}(\omega_\text{E})
	=
	-\frac{1}{|\omega_\text{E}|}
	\,.
\end{equation}
Inverse Fourier transforming (at zero temperature for simplicity) the Euclidean correlator to express it in position space, we find:
\begin{equation}
	\mathcal{G}^\text{E}_{\Delta=0}(\tau)
	=
	\frac{1}{2\pi}\left(
		2\gamma + \log|\tau|^2
	\right),
	\label{eq:GEtau}
\end{equation}
with $\gamma$ the Euler-Mascheroni constant.
Therefore, quantum averaging \eqref{eq:GEtau} is effectively computing the average value $\vev{\log|\tau|^2} = {-} \vev{\log \mathcal{G}^{\text{E}}_{\Delta=1}}$.
This quantity can be obtained by expanding the Euclidean correlator for generic $\Delta$, and identifying the logarithm as the order $\Delta$ contribution:
\begin{equation}
	\vev{\mathcal{G}^\text{E}_\Delta}
	=
	\vev{\left(\mathcal{G}^\text{E}_{\Delta=1}\right)^\Delta}
	=
	1 
	+ \Delta \vev{\log{\mathcal{G}^\text{E}_{\Delta=1}}}
	+ O(\Delta^2)
	\, .
\end{equation}
We then take:
\begin{equation}
	\vev{\mathcal{G}^\text{E}_{\Delta=0}(\tau)}
	=
	\frac{1}{2\pi}\left(
		2\gamma - \vev{\log{\mathcal{G}^\text{E}_{\Delta=1}}}
	\right)\, .
\end{equation}

With this interpretation of the $\Delta=0$ correlator, we use our results above to compute the Matsubara coefficients \eqref{eq:Matsubara_coef} in the limit $\Delta \to 0$.
Expanding $\vev{\mathcal{G}_\Delta^\text{E}(\tau_\text{E})}$ from \eqref{eq:GEDelta_tree} and \eqref{eq:Delta_loop} with $\mathcal{N}_\Delta = 1$, we find
\begin{equation}
\begin{split}
	\vev{\mathcal{G}_\Delta^\text{E}(\tau_\text{E})}
	=&
	1 
	+ \Delta\left(
		\log \frac{(\pi/\beta)^2}{\sin^2 \frac{\pi}{\beta}|\tau_\text{E}|}
		+ \right.\\
        &\left.\quad+\frac{\beta}{4\pi^2 C} \left[
			1 
			+ \frac{2\pi}{\beta}\frac{
				\beta/2 - |\tau_\text{E}|
			}{
				\tan \frac{\pi}{\beta}|\tau_\text{E}|
			}
			- \frac{\pi^2}{\beta^2}\frac{
				|\tau_\text{E}|(\beta - |\tau_\text{E}|)
			}{
				\sin^2\frac{\pi}{\beta}|\tau_\text{E}|
			}
		\right]
	\right) 
	+ O(\Delta^2)\,.
\end{split}
	\label{eq:expansion_GEDelta}
\end{equation}
This result agrees with another approach to the calculation, performed by expanding in $\Delta$ before the saddle-point approximation, which is presented in Appendix~\ref{app:log}.
Comparing with:
\begin{equation}
	\vev{\mathcal{G}_\Delta^\text{E}(\tau_\text{E})}
	=
	1 + \Delta \left(
		4\pi\gamma - 2\pi\vev{\mathcal{G}_{\Delta=0}^\text{E}(\tau_\text{E})}
	\right)
	+ O(\Delta^2)
	\,,
\end{equation}
we obtain:
\begin{equation}
\begin{split}
	\vev{\mathcal{G}_{\Delta=0}^\text{E}(\tau_\text{E})}
	=&
	2\gamma
	-\frac{1}{2\pi}\log \frac{(\pi/\beta)^2}{\sin^2 \frac{\pi}{\beta}|\tau_\text{E}|}\\
    &
	- \frac{\beta}{8\pi^3 C} \left[
		1 
		+ \frac{2\pi}{\beta}\frac{
			\beta/2 - |\tau_\text{E}|
		}{
			\tan \frac{\pi}{\beta}|\tau_\text{E}|
		}
		- \frac{\pi^2}{\beta^2}\frac{
			|\tau_\text{E}|(\beta - |\tau_\text{E}|)
		}{
			\sin^2\frac{\pi}{\beta}|\tau_\text{E}|
		}
	\right]
	.
\end{split}
	\label{eq:GE0}
\end{equation}
%
The retarded Green's function is obtained by analytically continuing the Matsubara coefficients of \eqref{eq:GE0}, in a similar fashion as for $\Delta = 1$.  
The end result reads:
\begin{equation}
	\vev{\mathcal{G}^\text{R}_{\Delta=0}(\omega)}
	=
	\frac{1}{\ii\omega}
	\left[
		1
		+
		\frac{\beta}{4\pi^2 C}\left(
			- 1
			+ \ii\frac{\beta\omega}{\pi}
			- \frac{\beta^2\omega^2}{2\pi^2}\psi'(-\ii\beta\omega/2\pi)
		\right)
		+ O\left(\frac{\beta^2}{C^2}\right)
	\right].
	\label{eq:GR0omega}
\end{equation}
Expanding as $\beta\omega \ll 1$, we find
\begin{equation}
	\vev{\mathcal{G}^\text{R}_{\Delta=0}(\omega)}
	=
	\frac{1}{\ii\omega}\left(
		1
		+ \frac{\beta}{4\pi^2 C}
		+ O(\beta^2/C^2,\beta\omega)
	\right)
	,
    \label{eq:vevG0}
\end{equation}
which is free of any real part in the regime of interest.

\section{Quantum-corrected shear retarded Green's functions \label{sec:quantumcorrshear}}

We are finally in a position to compute the effect of bulk quantum fluctuations on the shear retarded Green's function. In section \ref{sec:Matching}, we computed the non-local shear retarded Green's function
\begin{equation}
    G_{xy,xy}^{\text{R}}(\omega,k) 
    ={}
    \frac{s_0}{4\pi}
    \frac{
        \omega^2 
        \vev{\mathcal{G}_{\Delta=0}^{\text{R}}(\omega)}
    }{
        1 
        - {r_\e   \over 12 } 
        \vev{\mathcal{G}_{\Delta=0}^{\text{R}}(\omega)}
    k^2}
    \,,\quad 
    s_0
    =
    {2\pi L^2 \over  \kappa^2 r_\erm^2}
    \,.
  \label{eq:quantumcorrectedshearGR}
\end{equation}
Here we identified the IR operator as $\Delta=0$ as explained in section \ref{sec:mixdbdycond}, and $\vev{\mathcal{G}_{\Delta = 0}^{\text{R}}(\omega)}$ is given in \eqref{eq:vevG0}. After including quantum corrections, the retarded Green's function continue to feature a quadratically-dispersing pole
\begin{align}
    \ii\omega={}& 
    \vev{D_\perp} k^2
    \,,\qquad 
    \vev{D_\perp}
    =
    \frac{r_\e}{12}\left[
        1 
        + \frac{1}{4\pi^2 C T}
        + O\!\left(\frac1{(CT)^2}\right)
    \right]
    \,.
\end{align}
We analyze this in the two near-extremal regimes of interest, the hydrodynamic regime $k\lesssim k_\eq$ and the non-hydrodynamic regime $k_\eq\lesssim k\ll1/r_\e$.

\paragraph{Schwarzian corrections to shear hydrodynamics}\strut\\
In the hydrodynamic regime, \eqref{eq:quantumcorrectedshearGR} leads to two different ways to extract $\vev{\eta}$: either from the dispersion relation of the pole
\begin{equation}
    \vev{\eta}
    =
    \vev{D_\perp}\chi_{\pi\pi}
    \,,
\end{equation}
where to this order in matching the static susceptibility $\chi_{\pi\pi}$ \eqref{chiPiPiRN} does not receive quantum corrections;
or from the Kubo formula:
\begin{equation}
\label{eq:quantcorretaKuboDelta0}
    \vev{\eta} 
    = \frac{s_0}{4\pi}
    \lim_{\omega\to0}\omega^2
    \vev{ \mathcal{G}_{\Delta=0}^{\text{R}}(\omega)}
    \,.
\end{equation}
Using \eqref{eq:vevG0}, 
this gives in both cases
\begin{equation}
\label{eq:quantcorretaDelta0}
    \vev{\eta}
    =
    \frac{s_0}{4\pi}\left[
        1 
        + \frac{1}{4\pi^2 C T}
        + O\!\left(\frac1{(CT)^2}\right)
    \right]
    .
\end{equation}

On the other hand, setting $k=0$ from the start of the calculation, we identified in section \ref{sec:dirbdycond} the dimension of the corresponding IR as $\Delta=1$, and so
\begin{equation}
\label{eq:quantcorretaDelta1}
    \vev{\eta} 
    = 
    \frac{s_0}{4\pi}
    \lim_{\omega\to0}\frac1\omega 
    \vev{ \mathcal G_{\Delta=1}^{\text{R}}(\omega)}
    \,.
\end{equation}
Using \eqref{eq:vevG1}
leads to the same result as \eqref{eq:quantcorretaDelta0}. Hence both approaches agree with each other.

\paragraph{KSS bound}\strut\\
To leading order, we find that the KSS bound \eqref{KSSbound} is preserved by the sector of quantum corrections computed in this work:
\begin{equation}
    \frac{\vev{\eta}}{s_0}
    =
    \frac{1}{4\pi}\left[
        1 
        + \frac{1}{4\pi^2 C T}
        + O\!\left(\frac1{(CT)^2}\right)
    \right]
    .
\end{equation}
We recall that as discussed at the end of section \ref{sec:shearclholo}, the consistency of our leading order matching calculation requires to approximate the thermodynamics quantities by their extremal semi-classical value, given $T r_\e\ll1$. This implies that the near-extremal $O(CT)$  correction to $s_0$ is outside the regime of validity of our calculation, and so is the $O(\log CT)$ quantum correction from the Schwarzian \cite{iliesiu_statistical_2021}. To capture these dependencies, we would need to do that matching to the next order in $\epsilon$, where temperature corrections appear and would also affect the IR retarded Green's function, not just the thermodynamics. However, since this is now sensitive to irrelevant deformations away from AdS$_2\times\mathbb{T}^2$, it is unclear whether the quantum-averaged IR correlator can still be computed in the same way. In this respect, our work differs from the other recent results on the same topic \cite{Cremonini:2025yqe,Kanargias:2025vul,Nian:2025oei,PandoZayas:2025snm}.

\paragraph{Schwarzian corrections beyond hydrodynamics}\strut\\
The result \eqref{eq:quantumcorrectedshearGR} is valid for arbitrary $\omega/T$, so long as $1/C\lesssim\omega, T\ll 1/r_\e$, in particular in the non-hydrodynamic regime $\omega/T\gtrsim1$. Recall that classically, a $T=0$ quadratic mode $\ii\omega=D_\perp(T=0)k^2+O(k^4)$ was preserved in this limit. Once quantum corrections are included, the small $T$ limit of the quantum-corrected quadratic coefficient is not smooth and instead diverges like $1/(CT)$. Thus, we expect the limit $CT\to O(1)$ not to leave intact the classical gapless mode. However, its precise fate lies beyond our current analysis.

\section{Discussion}

In this work, we computed the impact of quantum fluctuations originating from the universal Schwarzian theory controlling the near-extremal AdS$_2$ near-horizon geometry of the Reissner-Nordstr\"om planar black hole.

There are many directions in which our current results may be extended.

\paragraph{Next order in matching:}

Serendipitously, we found the same correction to the shear viscosity, whether we use the Kubo formula where the dimension of the IR operator is $\Delta =1$, or the location of the diffusive pole, which is controlled instead by a $\Delta=0$ operator. We are not aware of a reason that would explain a priori why the correction turns out to be the same and whether this should persist for higher-order corrections. To evaluate these, it is necessary to go to higher-order in matching.

Since for the shear sector of Reissner-Nordstr\"om, the inner solution does not depend on wavenumber to $O(\epsilon)$ \cite{Gouteraux:2026}, it is likely possible to compute the effect of Schwarzian quantum fluctuations to that order as well. Doing so would let us keep the leading temperature-dependent corrections away from extremality to the entropy density and to the shear viscosity. After quantum averaging, the entropy density receives a logarithmic correction \cite{iliesiu_statistical_2021}, which other works \cite{Cremonini:2025yqe,Nian:2025oei,PandoZayas:2025snm,Kanargias:2025vul} have argued is responsible for a violation of the KSS bound.

We would also be able to compute the impact of quantum fluctuations on the spectrum of SL$(2,\mathbb{R})$ gapped modes, or the fate of non-analytic contributions to the classical dispersion relation of the gapless $T=0$ mode.

\paragraph{Other sectors:} There is no obstacle a priori to computing the Schwarzian bulk quantum corrections to other cases of interest:
\begin{itemize}
    \item We have only addressed the shear sector. The charge sector is straightforward to tackle, since the inner solution there also does not depend on $k$ at leading order in $\epsilon$.
    \item The longitudinal sector comprises energy fluctuations and is the most interesting, as this is where the Schwarzian action may be expected to have the strongest impact. However, it presents a difficulty: there, the inner solution depends on $k$ at order $O(\epsilon^0)$, meaning that the massive modes of the spatial torus cannot be decoupled and need to be quantized. This is reflected in the fact that the retarded Green's function evaluated in the spatially-extended Schwarzian theory carries the energy diffusion pole, as has been computed in spatially-extended SYK models \cite{Gu:2016oyy,Davison:2016ngz,Choi:2020tdj}.
\end{itemize}

\paragraph{Other systems:} There is no obstacle a priori to compute the Schwarzian bulk quantum corrections to other cases of interest:
\begin{itemize}
    \item The 5-dimensional AdS-Reissner-Nordstr\"om solution is dual to $\mathcal N=4$ super Yang-Mills at infinite coupling and infinite number of colors $N_\text{c}$. The shear viscosity is an important observable to characterize the Quark-Gluon-Plasma in QCD. 
    \item Other examples of spacetimes with an extremal AdS$_2\times \mathbb{R}^d$ which could be investigated include brane constructions \cite{Gushterov:2018spg}, the translation-symmetry breaking solutions of \cite{Bardoux:2012aw,andrade_simple_2014} or spacetimes charged under a higher-form symmetry \cite{Grozdanov:2018ewh,Davison:2025sze}.
    \item Spacetimes with a near-extremal geometry conformal to AdS$_2\times \mathbb{R}^d$ arise in top-down constructions and have also been extensively used in applications of gauge/gravity duality to the problem of strange metallicity, since they have an entropy density vanishing linearly with temperature and upon breaking of spatial translations, a resistivity linear in temperature \cite{Davison:2013txa}. Both of these are properties of the normal phase of high $T_\text{c}$ superconductors, and a holographic model capturing at least some of these aspects is appealing. More to the point of this work, this spacetime also retains classical $T=0$ gapless modes \cite{Davison:2013uha}. More work would be needed to understand the relation to the Schwarzian, if any, and whether any of the methods developed to compute correlation functions in the Schwarzian effective theory can also be applied there.
\end{itemize}

\paragraph{Role of boundary conditions:} There is a mounting body of evidence that IR gapless modes, whether hydrodynamic or not, arise when alternate boundary conditions must be imposed on a bulk field, \cite{Nickel:2010pr,Grozdanov:2018ewh,Davison:2022vqh,Ghosh:2020lel,He:2021jna,He:2022jnc,He:2022deg}. This seems like a phenomenon worth understanding more precisely.

\paragraph{Effective hydrodynamic theories and fluctuations:} The effect of fluctuations (stochastic or quantum) on hydrodynamics may be captured by turning to the Schwinger-Keldysh formalism for real-time thermal quantum field theories, see \cite{Liu:2018kfw} for a review of the formalism and \cite{Chen-Lin:2018kfl,Jain:2020zhu,Michailidis:2023mkd,Mullins:2023ott,Jain:2023obu,Jain:2020hcu} for recent work. More specifically, from \cite{Chen-Lin:2018kfl}, non-analyticities are generated in the retarded Green's function for hydrodynamic diffusion, leading to a splitting of the diffusive mode into two complex modes by the branch cut, and various modifications of the functional dependence of the retarded Green's function, including a priori independent shifts of the conductivity and of the diffusivity. The former is computed from the Kubo formula and the latter from the location of poles, suggesting that both determinations of transport coefficients may no longer lead to the same result. This is not the case for our $O(\epsilon^0)$ calculation, but could occur at $O(\epsilon)$.

\subsection*{Acknowledgments}
We thank Guillaume Bossard, Sean Hartnoll, Akash Jain, Alexandros Karnagias, Elias Kiritsis, Thomas Mertens, Sameer Murthy, Olga Papadoulaki, Achilleas Porfyriadis, Mikel Sanchez-Garitaonandia and especially Roberto  Emparan for discussions related to the topic of this work. B.~G.~ thanks the Simons Center for Geometry and Physics, Stony Brook University, where this project was conceived, and the Tata Institute for Theoretical Physics, where some of this work was carried out. The work of DMR was partially supported by STFC consolidated grant ST/X000664/1 and by the Simons Investigator award \#620869.

\appendix

\section{Review of shear hydrodynamics \label{app:shearhydro}}

The long-distance, late-time dynamics of fluids is described by hydrodynamics. This is tantamount to only keeping track of conserved operators protected by global symmetries of the state. For the case of relativistic fluids charged under a global U(1) symmetry, they are the stress-energy tensor $T^{\mu\nu}(t,x,y)$ and the current $J^\mu(t,x,y)$.\footnote{Lowercase greek indices run over the field theory spacetime coordinates $\mu,\nu=t,x^i$, while lowercase Latin indices $i,j$ run over the field theory spatial dimensions.} In this work, we only consider conformal fluids in two spatial dimensions (see \cite{Kovtun:2012rj} for a review), to which $3+1$-dimensional, asymptotically AdS black holes with a planar horizon are dual.

The invariance under spacetime translations and U(1) transformations gives the conservation of the stress-energy tensor and of the current as operator identities
\begin{equation}
    \label{conseq}
    \nabla_\mu T^{\mu\nu}=0\,,\quad \nabla_\mu J^\mu=0\,.
\end{equation}

The linear response of the sector transverse to the direction of the wavenumber $(T^{ty},T^{xy},J^y)$ is governed by the diffusion of transverse momentum. This means that the low frequency, low wavenumber behavior of the transverse retarded Green's functions of the state only exhibit a single pole
\begin{equation}
    \label{sheardisprel}
    \omega_\perp 
    = 
    - \ii D_\perp k^2 
    + O(k^4/k_c^4)
    \,.
\end{equation}
The leading behavior of the dispersion relation is quadratic in the wavenumber, and its coefficient, the diffusivity $D_\perp$, is expressed as
\begin{equation}
    D_\perp 
    = 
    \frac{\eta}{\chi_{\pi\pi}}
    \,,\quad 
    \chi_{\pi\pi}
    \underset{\textrm{Lorentz symmetry}}{=}
    \varepsilon + p
    \,.
\end{equation}
Two quantities appear in the expression above. The first is the momentum static susceptibility $\chi_{\pi^i\pi^j}\equiv\partial T^{ti}/\partial v^j|_{T,\mu}=\delta^{ij}\chi_{\pi\pi}$ (we are only considering isotropic states, so this quantity is diagonal), which for relativistic fluids is simply given by the enthalpy $\chi_{\pi\pi}=\varepsilon+p=sT+\mu\rho$. The second is the shear viscosity, which is a transport coefficient appearing in the constitutive relation of the stress-energy tensor at first order in gradient
\begin{equation}
    \label{Tconstrel}
    \vev{T^{\mu\nu}}
    = 
    (\varepsilon + p)u^\mu u^\nu
    + pg^{\mu\nu}
    - \eta \sigma^{\mu\nu}
    + O\left(\nabla^2\right)
    \,.
\end{equation}
Here $g_{\mu\nu}$ is the metric of the spacetime on which the fluid is considered, $u^\mu$ is the fluid velocity normalized such that $u^\mu u_\mu=-1$ and $\sigma_{\mu\nu}$ is the shear tensor, which is the transverse traceless combination of the gradient of the fluid velocity
\begin{equation}
    \sigma^{\mu\nu} 
    = P^{\mu\alpha}P^{\nu\beta}\left(
        \nabla_{\alpha}u_\beta
        + \nabla_{\beta}u_\alpha
    \right)
    - P^{\mu\nu}\nabla_\alpha u^\alpha
    \,,\quad 
    P^{\mu\nu}
    =
    g^{\mu\nu} + u^\mu u^\nu
    \,.
\end{equation}
$P^{\mu\nu}$ is the projector transverse to the fluid velocity.

There are higher-order in $k$ corrections to the dispersion relation \eqref{sheardisprel}. They come from higher-gradient corrections to the constitutive relations \eqref{Tconstrel}. They are suppressed by appropriate powers of $k_c$, which we define as the momentum scale at which hydrodynamics breaks down, i.e. the radius of convergence of the gradient expansion of the constitutive relation \cite{Withers:2018srf,Grozdanov:2019kge,Grozdanov:2019kge}. There is a corresponding frequency scale $\omega_c\equiv |\omega_\perp(k\to k_c)|$.

Hydrodynamics does not only predict the location of poles, but the actual expression of the retarded Green's function itself for frequencies and wavenumbers in the hydrodynamic regime. For the transverse sector, aligning the wavenumber along the $x$ direction without loss of generality,
\begin{equation}
    G^{\text{R}}_{ty,ty} 
    = 
    \frac{\eta k^2}{\ii\omega-D_\perp k^2}\,,
    \quad 
    G^{\text{R}}_{xy,xy} 
    = 
    \frac{\eta \omega^2}{\ii\omega-D_\perp k^2}
\end{equation}
where the retarded Green's function are related by the stress-energy tensor Ward identity of the theory \eqref{conseq}. Importantly for our purposes, these expressions offer two ways to compute the shear viscosity: either through the independent computation of the momentum static susceptibility and of the diffusivity coming from the denominator of the Green's functions, or through the Kubo relation coming from the numerator of the Green's functions
\begin{equation}
\label{etadefs}
    \eta =\left\{
        \begin{array}{l}
            \chi_{\pi\pi}D_\perp\\[3.5pt]
        -\underset{\omega\to0}{\mathrm{lim}}\,\dfrac1\omega\mathfrak{Im}\,G^{\text{R}}_{xy,xy}(\omega,k=0)
        \end{array}
    \right.
\end{equation}
A strong consistency condition from the hydrodynamic theory is that both of these expressions must agree.

\section{Detailed Next-to-leading Order Calculation}
\label{app:NLO}

In this appendix, we detail the next-to-leading order.
We want to evaluate \eqref{eq:GEDelta_saddle} up to order $O(x)$.
To be able to use Jordan's lemma and apply the residue theorem, we need to perform a Wick rotation.
Assuming $\tau_\text{E}>0$, analytic continuation yields $\tau_\text{E}\to \ii\tau + 0^+ $, and we have for the Wightman correlator
\begin{equation}
	\begin{split}
	\vev{\mathcal{G}_\Delta^-(\tau)}
	=&~
	\frac{\mathcal{N}_\Delta}{2\pi}
	\left(\frac{2\pi}{\beta}\right)^{2\Delta - 1}
	\int\dd \omega\,
	\e^{- \ii\omega(\tau - \ii 0^+)}
	\e^{\beta\omega/2}
	\frac{
		\Gamma(\Delta \pm \ii \beta\omega/2\pi)
	}{
		\Gamma(2\Delta)
	}\times
	\\
	&~
	\left(
		1 
		+ x \left[
			\frac{\left(m^{(1)}\right)^2}{8\pi^2}
			- \frac{(\beta\omega)^2}{32\pi^2}
			+ \frac{2\Delta - 1}{8\pi^2}
			- \frac{(\beta\omega)^2}{16\pi^4}
			\mathfrak{Re}\,\psi'\!\left[
				\Delta + \ii \beta\omega/2\pi
			\right]
		\right]
		+ O(x^2)
	\right)
	\\
	\equiv&~
	\frac{1}{2\pi}\int\dd \omega\,
	\e^{- \ii\omega(\tau - \ii 0^+)}
	\vev{\mathcal{G}_\Delta^-(\omega)}
	\,,
	\end{split}
\end{equation}
where $m^{(1)}$ is defined in \eqref{eq:saddle_Delta}.
Poles of the integrand are still located at 
$\omega^\pm_n = \pm 2\ii\pi(\Delta + n)/\beta$.

Assuming $\tau < 0$, the contour is closed in the upper half complex $\omega$ plane, whereas for $\tau > 0$ it is closed in the lower half plane.
In the following expressions, and the remaining of this appendix, the top sign corresponds to $\tau<0$, and the bottom sign to $\tau>0$.
For these two cases, the residue theorem implies:
\begin{equation}
	\vev{\mathcal{G}_\Delta^-(\tau)}
	=
	\pm \ii \sum_{n\in\mathbb{N}}
	\underset{\omega = \omega^\pm_n}{\mathrm{Res}}
	\!\left(
		\e^{- \ii\omega(\tau - \ii 0^+)}
		\vev{\mathcal{G}_\Delta^-(\omega)}
	\right)
	=
	\pm \ii \sum_{n\in\mathbb{N}}\left(
		R_n^{\pm,(0)} + x\, R_n^{\pm,(1)} + O(x^2)
	\right)
	,
\end{equation}
where we define order by order in $x$ the contributions to the residues that lie along respectively the positive and negative imaginary axis, $R_n^{+,(\text{order})}$ and $R_n^{-,(\text{order})}$.
We find for these contributions:
\begin{subequations}
\begin{align}
	R_n^{\pm,(0)}
	=&~
	\mp\ii\mathcal{N}_\Delta
	\left(
		\pm\frac{2\ii\pi}{\beta}\e^{\pm\pi (\tau-\ii 0^+)/\beta}
	\right)^{2\Delta}\frac{(2\Delta)_n}{n!}
	(\e^{\pm 2\pi (\tau-\ii 0^+)/\beta})^n
	\\	
	\frac{R_n^{\pm,(1)}}{R_n^{\pm,(0)}}
	=&~
	\Delta(2\Delta + 1)\frac{1}{4\pi^2}
	\pm (2\Delta + 1)\frac{\ii(\beta/2 - \ii\tau^-)}{2\pi\beta}
	(\Delta + n)
	+ \frac{\ii\tau^-(\beta - \ii\tau^-)}{2\beta^2}
	(\Delta + n)^2
	\,,
\end{align}
\end{subequations}
with $\tau^- = \tau - \ii 0^+$.
Irrespective of the sign of $\tau$, summing over the residues gives:
\begin{subequations}
\begin{align}
	\vev{\mathcal{G}_\Delta^-(\tau)}_\text{tree}
	=&~
	\mathcal{N}_\Delta
	\left(
		\frac{\pi}{\beta}
		\frac{1}{\ii\sinh\frac{\pi}{\beta}(\tau - \ii 0^+)}
	\right)
	\\	
	\begin{split}
	\frac{
		\vev{\mathcal{G}_\Delta^-(\tau)}_\text{one-loop}
	}{
		\vev{\mathcal{G}_\Delta^-(\tau)}_\text{tree}
	}
	=&~
	\Delta\left[
		(2\Delta + 1)\frac{1}{4\pi^2}
		+ (2\Delta + 1)
		\frac{1}{2\pi\beta}
		\frac{
			\beta/2 - \ii \tau^-
		}{
			\tan(\pi\tau^-/\beta)
		}
		+ \frac{1}{4\beta^2}\ii\tau^-(\beta-\ii\tau^-)
	\right.
	\\
	&~\left.
		(\Delta + 1)
		\frac{1}{4\beta^2}
		\frac{
			\ii\tau^-(\beta - \ii\tau^-)
		}{
			\sinh(\pi\tau^-/\beta)^2
		}
		+ \Delta \frac{1}{4\beta^2}\ii\tau^-(\beta - \ii\tau^-)
		(1 - \coth(\pi\tau^-/\beta)^2)
	\right]
	.
	\end{split}
	\label{eq:Delta_loop_lorentzian}
\end{align}
\end{subequations}
Analytically continuating $\tau^- \to \tau_\text{E}>0$ of \eqref{eq:Delta_loop_lorentzian}, one finds \eqref{eq:Delta_loop}.
Repeating the analysis for $\tau_\text{E}<0$, one concludes about the absolute value in \eqref{eq:Delta_loop}.

\section{Another Approach to the Logarithmic Correlator \label{app:log}}

In order to extract $\vev{\log{\mathcal{G}^\text{E}_{\Delta=1}}}$ out of \eqref{eq:vev_2pt}, we can introduce a sequence of functions converging to a Dirac delta distribution as $\Delta\to 0^+$ as:
\begin{equation}
	\eta_\Delta(p_1 - p_2)
	=
	\frac{1}{2\pi}\frac{\Gamma(\Delta \pm \ii (p_1 - p_2))}{\Gamma(2\Delta)}
	\underset{\Delta\to 0^+}=
	\delta(p_1 - p_2) + \Delta\frac{1}{(p_1-p_2)\sinh[\pi(p_1 - p_2)]}
	+ O(\Delta^2)
	\,.
\end{equation}
Expanding the quantum-corrected Euclidean Green's function, one finds:
\begin{equation}
	\vev{\mathcal{G}_\Delta^\text{E}(\tau)}
	\underset{\Delta\to 0^+}=
	1
	+ \Delta\left( I_1 + I_2(\tau) \right)
	+ O(\Delta^2)
	=
	1 
	+ \Delta\vev{\log{\mathcal{G}^\text{E}_{\Delta = 1}}}
	+ O(\Delta^2)
	\,,
\end{equation}
with:
\begin{subeq}
	I_1
	=&~
	\frac{C}{2\pi^2}\frac{\e^{S_0}}{\mathcal{Z}(\beta)}
	\int\dd E\,\e^{-\beta E}\sinh[2\pi\sqrt{2CE}]\left(
		2\mathfrak{Re}\,\psi[2\ii \sqrt{2CE}]
		- 2\log{2C}
	\right),\label{eq:I1}%
	\\
	I_2(\tau)
	=&~
	\frac{C}{2\pi^2}\frac{\e^{S_0}}{\mathcal{Z}(\beta)}
	\int\dd\omega\dd M\,
	\frac{\e^{(\beta/2-|\tau|)\omega}}{\omega}
	\frac{
		\e^{-\beta M}\sinh[2\pi\sqrt{2C(M\pm\omega/2)}]
	}{
		\sinh[\pi\sqrt{2C}(\sqrt{M+\omega/2}\pm\sqrt{M-\omega/2})]
	}\,.\label{eq:I2}%
\end{subeq}

Note that the saddle-point analysis to evaluate the $M$ integral inside $I_2$ in \eqref{eq:I2} is identical to the defined in \eqref{eq:saddle-eq} by setting $\Delta=1$.
This means that $I_2$ can be written using a saddle-point approximation as:
\begin{equation}
	I_2(\tau)
	=
	\int_{-\infty}^{+\infty}\dd w\,
	\e^{w(\pi-u)/2\pi}\frac{1}{w\sinh{w/2}}\left(
		1
		+ \frac{x}{16\pi^2}\frac{w^2}{\sinh(w/2)^2}
		+ O(x^2)
	\right),
	\label{eq:I2x}
\end{equation}
with $u=2\pi|\tau|/\beta$, $w=\beta\omega$ and $x=\beta/C$.

Evaluation the $I_1$ integral in \eqref{eq:I1} can also be done through a saddle point approximation, defining a function $F_1$:
\begin{subeq}
	I_1 
	=&~
	- 2\log{2C} 
	+ \frac{C}{\pi^2}\frac{\e^{S_0}}{\mathcal{Z}(\beta)}\int\dd E\,
	\e^{-F_1(E;C)}\,,
	\\
	F_1(E;C)
	=&~
	\beta E
	-
	\log{\sinh[2\pi\sqrt{2CE}]}
	- \log{\mathfrak{Re}\,\psi[2\ii\sqrt{2CE}]}\,.
\end{subeq}
Looking for the minimum, we do the change of variable $E=C \varepsilon/\beta^2$ and define $x=\beta/C$ as our parameter for the perturbative expansion around the semi-classical saddle.
The equation to solve is then:
\begin{equation}
	0
	\overset{!}=
	\beta^{-1}\partial_M F_1
	=
	1
	-
	\sqrt{\frac{2\pi^2}{\varepsilon}}\coth[2\pi\sqrt{2\varepsilon}/x]
	+ \sqrt{\frac{2}{\varepsilon}}
	\frac{
		\mathfrak{Im}\,\psi'[2\ii\sqrt{2\varepsilon}/x]
	}{
		\mathfrak{Re}\,\psi[2\ii\sqrt{2\varepsilon}/x]
	}\,.
\end{equation}
Expanding in $x\ll 1$, defining $\varepsilon = \varepsilon^{0} + x\, \varepsilon^{(1)} + x^2\,\varepsilon^{(2)} + \ldots\,$, we find:
\begin{equation}
	0
	\overset{!}=
	1 - \sqrt{\frac{2\pi^2}{\varepsilon^{(0)}}}
	+ \frac{x}{\varepsilon^{(0)}} \left(
		\frac{\pi \varepsilon^{(1)}}{\sqrt{2\varepsilon^{(0)}}}
		- \frac{1}{2\log{(2\sqrt{2\varepsilon^{(0)}}/x)}}
	\right)
	+ O(x^2)\,,
\end{equation}
which yields:
\begin{equation}
	\varepsilon^{(0)}
	=
	2\pi^2\,,
	\qquad
	\varepsilon^{(1)}
	=
	\frac{1}{\log(4\pi/x)}\,.
\end{equation}
We find that expansion is non-analytic in $x$.
Computing the width, we obtain:
\begin{equation}
	\sqrt{\frac{2\pi}{\partial_M^2 F_1(E^\star)}}
	=
	\frac{1}{C}\left(\frac{2\pi C}{\beta}\right)^3
	\left(
		1 + \frac{x}{8\pi^2} \frac{1 - \log(4\pi/x)}{\log(4\pi/x)^2}
	\right)
	+ 
	O(x^2)\,.
\end{equation}
The saddle point approximation therefore yields:
\begin{equation}
	\begin{split}	
		I_1
		=&~
		- 2\log{2C} 
		+ \frac{C}{\pi^2}\frac{\e^{S_0}}{\mathcal{Z}(\beta)}
		\sqrt{\frac{2\pi}{\partial_M^2 F_1(E^\star;C)}}
		\e^{-F_1(E^\star;C)}
		\\
		=&~
		2\log(2\pi/\beta) + \frac{x}{4\pi^2} + O(x^2)\,.
	\end{split}
	\label{eq:I1sol}
\end{equation}

Having a look at \eqref{eq:I2x}, we see that the integral is manifestly non-integrable, the integrand exhibiting a double pole at $w=0$.
In order to assign a finite value to this integral, we use the following procedure:
\begin{enumerate}
	\item differentiate with respect to $\tau$, which amounts to multiply the integrand with a factor of $w$, turning the double pole at $w=0$ into a simple pole,
	\item evaluate the principal value of this integral, which can be done since the integration path over the real line only crosses a simple pole,
	\item integrate the result to assign, up to a constant, a finite value to Eq.~\eqref{eq:I2x}.
\end{enumerate}

Differentiating $I_2$ yields for the principal value to evaluate:
\begin{equation}
	\begin{split}
	-\mathrm{sgn}(\tau)\beta \partial_\tau I_2
	=&~
	\mathrm{p.v.}\int_{-\infty}^{+\infty}
	\dd w\, \frac{\e^{w(\pi - u)/2\pi}}{\sinh[w/2]}\left(
		1 
		+ \frac{x}{16\pi^2}\frac{w^2}{\sinh[w/2]^2}
		+ O(x^2)
	\right)
	\\
	=&~
	2\int_{0}^{+\infty}
	\dd w\, \frac{\sinh[w(\pi - u)/2\pi]}{\sinh[w/2]}\left(
		1 
		+ \frac{x}{16\pi^2}\frac{w^2}{\sinh[w/2]^2}
		+ O(x^2)
	\right)
	\\
	=&~
	\frac{\pi}{\tan[w/2]}
	+ \frac{x}{16\pi}\left(
		\frac{(2\pi-u)u - 4(\pi - u)}{\sin[u/2]^2}
		- \frac{4(\pi - u)}{\tan[u/2]}
	\right)
	+ O(x^2)
	.
	\end{split}
\end{equation}
Then, integrating \wrt $\tau$, we find $I_2$ up to a constant:
\begin{equation}
	I_2
	=
	- 2\log{|\!\sin{[u/2]}|^2}
	+ \frac{x}{16\pi^2}\left[
		\frac{u(u-2\pi)}{\sin[u/2]^2}
		+ \frac{4(\pi - u)}{\tan[u/2]}
	\right]
	+ \mathrm{cst}
	+ O(x^2)
	\,.
	\label{eq:I2sol}
\end{equation}
Remember that $u = 2\pi|\tau|/\beta$.

Putting everything together, namely \eqref{eq:I1sol} and \eqref{eq:I2sol}, we have:
\begin{equation}
    \begin{split}
        \vev{\log{\mathcal{G}^\text{E}_{\Delta=1}}}
        =&~
        I_1 + I_2
        \\
        =&~
        - \log{\frac{\beta^2}{4\pi^2}|\!\sin{[u/2]}|^2}
        + \frac{x}{16\pi^2}\left[
		      4 
		      + \frac{4(\pi-u)}{\tan[u/2]} 
		      + \frac{u(u-2\pi)}{\sin[u/2]^2}
        \right]
        + \mathrm{cst}
        + O(x^2)
        \,,
    \end{split}
\end{equation}
which matches (up to a constant) with \eqref{eq:expansion_GEDelta}.

\section{Fourier-transforming $\mathcal{G}^{\text{E}}_\Delta$ \label{app:FT}}

For generic $\Delta$, the Euclidean space Green's function reads (as a function of $\tau_{\text{E}} = \tau_1^\text{E}-\tau_2^\text{E}$):
\begin{align}
    \mathcal{G}^\text{E}_\Delta(\tau_\text{E}) 
    ={}& 
    \mathcal{N}_\Delta \left(\frac{\pi}{\beta}
    {1 \over \sin( {\pi \over \beta} |\tau_\text{E}|)}
    \right)^{2\Delta}
    \,.
\end{align}
The (Euclidean) frequency space Green's function is then:
\begin{equation}
    \mathcal{G}^\text{E}_\Delta(\omega_n^\text{E}) 
    = 
    \int_0^\beta \dd \tau_\text{E}\, 
    \e^{\ii \omega_n^\text{E} \tau_\text{E}} 
    \mathcal{G}^{\text{E}}_\Delta(\tau_\text{E}) 
    \,, 
    \label{eq:FTapp-GE-omegan-def}
\end{equation}
where $\omega_n^\text{E} = 2\pi n/\beta$ are the bosonic Matsubara frequencies.
To evaluate this integral, we first use the identity (assuming $\Delta >0$)
\begin{align}
  {1 \over A^\Delta} ={}& {1 \over \Gamma(\Delta)} \int_0^\infty \dd \lambda \lambda^{\Delta - 1} e^{{-} \lambda A}\, , 
\end{align}
perform the integral over $\tau_\text{E}$ in terms of modified Bessel
functions, and finally perform the $\lambda$ integral to obtain%
\footnote{Strictly speaking, the integral over $\lambda$ only converges for $0< \Delta < \frac{1}{2}$. The result for $\Delta >\frac{1}{2}$ is then obtained via analytic continuation. Alternatively, one can explicitly introduce a cutoff $\epsilon$ in \eqref{eq:FTapp-GE-omegan-def} by integrating from $\tau_\text{E}=\epsilon$ to $\tau_\text{E}=\beta-\epsilon$ and discarding the pieces of the resulting integral that diverge as $\epsilon\to 0$; this yields the same result as analytic continuation from $\Delta < \frac{1}{2}$ result.}:
\begin{align}
    \mathcal{G}^\text{E}_\Delta(\omega_n^\text{E}) 
    ={}& 
    \mathcal{N}_\Delta\sqrt{\pi} \left(\frac{\pi}{\beta}\right)^{2\Delta -1} 
    {\Gamma\left({1 \over 2} - \Delta \right) 
    \over 
    \Gamma(\Delta)} 
    {\Gamma\left(
        \Delta + {\beta\vert \omega_n^\text{E} \vert \over 2\pi} 
    \right) 
    \over 
    \Gamma\left(
        1 - \Delta 
        + {\beta\vert \omega_n^\text{E} \vert \over 2\pi} 
    \right)}
    \, .
\end{align}

We are principally interested in the $\Delta \to 0$ case, where the
position space correlator exhibits logarithmic behavior:
\begin{align}
    \mathcal{N}_\Delta^{-1}\mathcal{G}_\Delta^\text{E}(\tau_\text{E}) 
    ={}& 
    1 
    - \Delta \log \left[
        {\beta^2 \over \pi^2} 
        \sin^2\left({\pi \tau_\text{E} \over \beta} \right) 
    \right] 
    + O(\Delta^2)
    \,.
\end{align}
In this limit, the Fourier transform reduces to\footnote{The $n=0$
case can be found by first setting $n=0$ in the generic $\Delta$
result and then sending $\Delta \to 0$, or by directly evaluating
\begin{align}
    \mathcal{N}_\Delta^{-1}\int_0^\beta \dd \tau_\text{E}\, 
    \mathcal{G}^\text{E}(\tau_\text{E}) 
    ={}& 
    \beta 
    - 2 \Delta \int_0^\beta \dd \tau_\text{E}\, 
    \log \left[
        {\beta \over \pi} 
        \sin \left({\pi \tau_\text{E} \over \beta} \right) 
    \right] 
    + O(\Delta^2) 
    = 
    \beta 
    - 2\beta \Delta \log \left({\beta \over 2\pi} \right) 
    + O(\Delta^2)
    \,.
\end{align}
}
\begin{align}
    \mathcal{N}_\Delta^{-1}\mathcal{G}_\Delta^\text{E}(\omega_n^\text{E}) 
    ={}&
    \begin{cases}
        {2 \pi \Delta \over \vert \omega_n^\text{E} \vert} 
        + O(\Delta^2) 
        & 
        n \neq 0\,, 
        \\
        \beta 
        - 2\beta \Delta \log \left({\beta \over 2\pi} \right) 
        + O(\Delta^2) 
        & 
        n = 0\,. 
    \end{cases}
\end{align}
As a check on this result, we can explicitly perform the Matsubara sum to recover the position space result:
\begin{align}
    \mathcal{N}_\Delta^{-1}\mathcal{G}^\text{E}_{\Delta}(\tau_\text{E}) 
    ={}& 
    {1 \over \beta} \sum_n e^{-\ii \omega_n^\text{E} \tau_\text{E}} 
    \mathcal{G}^\text{E}_\Delta(\omega_n) 
    = 
    {\mathcal{G}^\text{E}_\Delta(\omega_0^\text{E}) 
    \over 
    \beta} 
    + {4\pi \Delta \over \beta} 
    \sum_{n=1}^\infty 
    {\cos (\omega_n^\text{E} \tau_\text{E} )
    \over 
    \omega_n^\text{E}} 
    + O(\Delta^2)
    \nonumber 
    \\
    ={}& 
    1 
    - \Delta \log \left[ 
        {\beta^2 \over \pi^2} \sin^2\left( 
            {\pi \tau_\text{E}\over \beta}
        \right) 
    \right] 
    + O(\Delta^2)
    \,.
\end{align}

To obtain the retarded Green's function, we analytically continue $\mathcal{G}^\text{E}_\Delta(\omega_n^\text{E})$ for $n>0$ to the real axis, $\ii \omega_n^\text{E} \to \omega+ \ii \epsilon$, finding:
\begin{align}
    \mathcal{N}_\Delta^{-1}\mathcal{G}_\Delta^\text{R}(\omega) 
    ={}& 
    {-}{2\pi \Delta \over \ii \omega} 
    + O(\Delta^2)
    \, .
\end{align}

\clearpage

\bibliographystyle{JHEP}
\bibliography{quantum_fluctuations.bib}

\end{document}